\documentclass[letterpaper]{article}
\usepackage{amsmath}
\usepackage{amssymb}
\usepackage{overcite}
\usepackage{graphicx}

\catcode`\@=11
\long\def\@makefntext#1{
\protect\noindent \hbox to 3.2pt {\hskip-.9pt  
$^{{\eightrm\@thefnmark}}$\hfil}#1\hfill}		

\def\thefootnote{\fnsymbol{footnote}}
\def\@makefnmark{\hbox to 0pt{$^{\@thefnmark}$\hss}}	
	
\def\ps@myheadings{\let\@mkboth\@gobbletwo
\def\@oddhead{\hbox{}
\rightmark\hfil\eightrm\thepage}   
\def\@oddfoot{}\def\@evenhead{\eightrm\thepage\hfil
\leftmark\hbox{}}\def\@evenfoot{}
\def\sectionmark##1{}\def\subsectionmark##1{}}

\oddsidemargin=\evensidemargin
\renewcommand{\thefootnote}{\fnsymbol{footnote}}

\newcounter{sectionc}
\newcounter{subsectionc}[sectionc]
\newcounter{subsubsectionc}[subsectionc]
\renewcommand{\thesectionc}{\Roman{sectionc}}
\renewcommand{\thesubsectionc}{\Alph{subsectionc}}
\renewcommand{\thesubsubsectionc}
	{\Roman{sectionc}.\arabic{subsectionc}.\arabic{subsubsectionc}}

\renewcommand{\section}[1] {\vspace{12pt}
        \refstepcounter{sectionc}  
	\begin{center}{\tenbf\thesectionc.  #1}\end{center}\par\vspace{5pt}} 

\renewcommand{\subsection}[1] {\vspace{12pt}
	\refstepcounter{subsectionc}
	\begin{center}{\thesubsectionc. {\kern1pt \tenit
		#1}}\end{center}\par\vspace{5pt}} 

\renewcommand{\subsubsection}[1] {\vspace{12pt}
        \refstepcounter{subsubsectionc}  
	\noindent{\tenrm\thesubsubsectionc. {\kern1pt \tenit
	#1}}\par\vspace{5pt}}  

\newcommand{\nonumsection}[1] {\vspace{12pt}\noindent{\tenbf #1}
	\par\vspace{5pt}}

\newcounter{appendixc}
\newcounter{subappendixc}[appendixc]
\newcounter{subsubappendixc}[subappendixc]
\renewcommand{\thesubappendixc}{\Alph{appendixc}.\arabic{subappendixc}}
\renewcommand{\thesubsubappendixc}
	{\Alph{appendixc}.\arabic{subappendixc}.\arabic{subsubappendixc}}

\renewcommand{\appendix}[1] {\vspace{12pt}
        \setcounter{figure}{0}
        \setcounter{table}{0}
        \setcounter{lemma}{0}
        \setcounter{theorem}{0}
        \setcounter{corollary}{0}
        \setcounter{definition}{0}
        \setcounter{equation}{0}
        \renewcommand{\thefigure}{\Alph{appendixc}.\arabic{figure}}
        \renewcommand{\thetable}{\Alph{appendixc}.\arabic{table}}
        \renewcommand{\theappendixc}{\Alph{appendixc}}
        \renewcommand{\thelemma}{\Alph{appendixc}.\arabic{lemma}}
        \renewcommand{\thetheorem}{\Alph{appendixc}.\arabic{theorem}}
        \renewcommand{\thedefinition}{\Alph{appendixc}.\arabic{definition}}
        \renewcommand{\thecorollary}{\Alph{appendixc}.\arabic{corollary}}
        \renewcommand{\theequation}{\Alph{appendixc}\arabic{equation}}
        \refstepcounter{appendixc}
	\noindent{\tenbf Appendix \theappendixc. #1}\par\vspace{5pt}}
\newcommand{\subappendix}[1] {\vspace{12pt}
        \refstepcounter{subappendixc}
        \noindent{\bf Appendix \thesubappendixc. {\kern1pt \bf #1}}
	\par\vspace{5pt}}
\newcommand{\subsubappendix}[1] {\vspace{12pt}
        \refstepcounter{subsubappendixc}
        \noindent{\rm Appendix \thesubsubappendixc. {\kern1pt \tenit #1}}
	\par\vspace{5pt}}

\newcommand{\textlineskip}{\baselineskip=13pt}
\newcommand{\smalllineskip}{\baselineskip=10pt}

\def\eightcirc{
\begin{picture}(0,0)
\put(4.4,1.8){\circle{6.5}}
\end{picture}}
\def\eightcopyright{\eightcirc\kern2.7pt\hbox{\eightrm c}}

\def\abstracts#1#2#3{{
	\centering{\begin{minipage}{4.5in}\baselineskip=20pt
	\parindent=0pt #1\par 
	\parindent=15pt #2\par
	\parindent=15pt #3
	\end{minipage}}\par}} 

\def\keywords#1{{
	\centering{\begin{minipage}{4.5in}\baselineskip=10pt\footnotesize
	{\footnotesize PACS}\/ #1
	 \end{minipage}}\par}}

\renewenvironment{thebibliography}[1]
        {\frenchspacing
	 \ninerm\baselineskip=11pt
         \begin{list}{$^{\arabic{enumi}}$}
        {\usecounter{enumi}\setlength{\parsep}{0pt}     
	 \setlength{\leftmargin 12.7pt}{\rightmargin 0pt} 
         \setlength{\itemsep}{5pt} \settowidth
	{\labelwidth}{$^{#1}$}\sloppy}}{\end{list}}

\newcounter{itemlistc}
\newcounter{romanlistc}
\newcounter{alphlistc}
\newcounter{arabiclistc}

\newcommand{\fcaption}[1]{
        \refstepcounter{figure}
        \setbox\@tempboxa = \hbox{\footnotesize Fig.~\thefigure. #1}
        \ifdim \wd\@tempboxa > 5in
           {\begin{center}
        \parbox{5in}{\footnotesize\smalllineskip Fig.~\thefigure. #1}
            \end{center}}
        \else
             {\begin{center}
             {\footnotesize Fig.~\thefigure. #1}
              \end{center}}
        \fi}

\renewcommand{\thetable}{\Roman{table}}
\newcommand{\tcaption}[1]{
        \refstepcounter{table}
        \setbox\@tempboxa = \hbox{\footnotesize Table~\thetable. #1}
        \ifdim \wd\@tempboxa > 5in
           {\begin{center}
        \parbox{5in}{\footnotesize\smalllineskip Table~\thetable. #1}
            \end{center}}
        \else
             {\begin{center}
             {\footnotesize Table~\thetable. #1}
              \end{center}}
        \fi}

\def\pmb#1{\setbox0=\hbox{#1}
	\kern-.025em\copy0\kern-\wd0
	\kern.05em\copy0\kern-\wd0
	\kern-.025em\raise.0433em\box0}

\def\fnt#1#2{\footnotetext{\kern-.3em
	{$^{\mbox{\scriptsize #1}}$}{#2}}}
\def\fpage#1{\begingroup
\voffset=.3in
\thispagestyle{empty}\begin{table}[b]\centerline{\footnotesize #1}
	\end{table}\endgroup}

\font\tenrm=cmr10
\font\tenit=cmti10 
\font\tenbf=cmbx10

\font\ninerm=cmr9

\font\eightrm=cmr8

\def\qed{\hbox{${\vcenter{\vbox{			
   \hrule height 0.4pt\hbox{\vrule width 0.4pt height 6pt
   \kern5pt\vrule width 0.4pt}\hrule height 0.4pt}}}$}}

\renewcommand{\thefootnote}{\fnsymbol{footnote}}	

\newcommand{\dd}{\mbox{d}}
\newcommand{\e}{\mbox{e}}

\newcommand{\di}[2]{\frac{\partial #1}{\partial #2}}
\newcommand{\ddi}[2]{\frac{\partial^2 #1}{\partial {#2}^2}}
\newcommand{\diff}[2]{\frac{\dd #1}{\dd #2}}
\newcommand{\ddiff}[2]{\frac{\dd^2 #1}{\dd {#2}^2}}
\newcommand{\G}[2]{G_{#1}^{\hspace{5pt} #2}}

\newcommand{\T}[2]{T_{#1}^{\hspace{5pt} #2}}

\newlength{\InitialBLS}
\newlength{\singlespace}
\newlength{\doublespace}
\begin{document}


\normalsize\textlineskip
\thispagestyle{empty}
\setcounter{page}{1}


\vspace*{0.88truein}
%
%
\fpage{1}
\centerline{\Large \bf Stability of Gravitational and}
\vspace*{0.035truein}
\centerline{\Large \bf  Electromagnetic Geons}
%
%
\vspace*{0.37truein}
\centerline{
G. P. Perry 
and
F. I. Cooperstock} 
%
\vspace*{0.015truein}
\centerline{
\it Department of Physics and Astronomy,
University of Victoria,}
\baselineskip=10pt
\centerline{
\it P.O. Box 3055, Victoria, B.C. V8W 3P6,
Canada}
%
\vspace*{0.225truein}

%
%
\vspace*{0.21truein}
 
\abstracts{Recent work on gravitational geons is extended to examine
the stability properties of gravitational and electromagnetic geon
constructs. All types of geons must possess the property of
regularity, self-consistency and quasi-stability on a time-scale much
longer than the period of the comprising waves. Standard perturbation
theory, modified to accommodate time-averaged fields, is used to test
the requirement of quasi-stability. It is found that the modified
perturbation theory results in an internal inconsistency. The
time-scale of evolution is found to be of the same order in magnitude
as the period of the comprising waves. This contradicts the
requirement of slow evolution. Thus not all of the requirements for
the existence of electromagnetic or gravitational geons are met though
perturbation theory. From this result it cannot be concluded that an
electromagnetic or a gravitational geon is a viable entity. The
broader implications of the result are discussed with particular
reference to the problem of gravitational energy.}{}{}

\vspace*{10pt}

%
%

\newpage
%
%


\vspace*{1pt}\textlineskip	
\vspace*{1pt}\setlength{\baselineskip}{\doublespace}
\section{Introduction}\label{introduction}	
\vspace*{-0.5pt}
\noindent
The examination of the basic properties of the gravitational field as
compared to other physical fields has concentrated around the
recently revived study of gravitational geons. The concept of a
structure comprised of electromagnetic waves held together by its own
gravitational attraction was first conceived by Wheeler
\cite{Wheeler55}. The extension of this idea using only gravitational
waves was first studied by Brill and Hartle\cite{BrillHartle}. Their
approach was to consider a strongly curved static `background
geometry' $\gamma_{\mu \nu}$ on top of which a small ripple $h_{\mu
\nu}$ resided, satisfying a linear wave equation. The wave frequency
was assumed to be so high as to create a sufficiently large effective
energy density which served as the source of the background
$\gamma_{\mu\nu}$, taken to be spherically symmetric on a
time-average. They claimed to have found a solution with a flat-space
spherical interior, a Schwarzschild exterior and a thin shell
separation meant to be created by high-frequency gravitational
waves. With the mass $M$ identified from the exterior metric, there
would follow an unambiguous realization of the gravitational geon as
described above.  It has since been
shown\cite{geonletter,geonstanford,geonpaper} that the Brill and
Hartle model does not implement the properties of high-frequency
waves, nor can the space-time be taken as singularity-free.

It was proposed by Cooperstock, Faraoni and Perry
\cite{geonletter,geonstanford,geonpaper} (henceforth referred to as
CFP) that a satisfactory gravitational geon model must be constructed
and solved in a manner similar to that of Wheeler's\cite{Wheeler55}
electromagnetic geon model. Such a model necessarily requires firstly
that the Einstein field equations be solved in a self-consistent
manner while satisfying the regularity conditions. Secondly, it is
required that the configuration represented by the metric
$\gamma_{\mu\nu}$ be quasi-stable over a time-scale much larger than
the typical period of its gravitational wave constituents (i.e.\ a
geon must maintain its bounded configuration for a sufficient length
of time, for otherwise it would not be possible to attribute a
structural form to the gravitational geon). Thirdly, it is required
that the gravitational field becomes asymptotically flat at spatial
infinity.  Thus a {\em gravitational (electromagnetic) geon} is a
bounded configuration of gravitational (electromagnetic) waves whose
gravity is sufficiently strong to keep them confined on a time-scale
long compared to the characteristic composing wave period. For the
gravitational case, it is required that no matter or fields other than
the gravitational field be present.

Through a series of papers, it was established by Anderson and
Brill\cite{AndersonBrill} and by CFP that in the high-frequency
approximation for a static background metric, the gravitational geon
problem and the electromagnetic geon problem are governed by the same
set of ordinary differential equations (ODEs) and boundary conditions.
These equations are satisfactory for considering the regularity and
self-consistency aspects of the geon problem but not the evolution in
time. Any solutions to these equations are necessarily equilibrium
solutions since the background metric is assumed static. Admissible
equilibrium solutions satisfying the boundary conditions have been
shown to exist\cite{AndersonBrill}. This paper provides an expanded
study of the gravitational and electromagnetic geon problem with
particular emphasis upon the dynamic evolution of geon constructs.
Section~\ref{Ch9} re-examines the numerical solutions to the equations
studied in Ref.~\citen{Wheeler55} and extends the analysis to obtain
information on the ``stability'' of these solutions with respect to
perturbations of the amplitude eigenvalues. The word ``stability''
used in Sec.~\ref{Ch9} should be viewed in the context of the boundary
conditions of a spatial variable, not a dynamic time variable.  Both
the analytic behaviour at spatial infinity and the aforementioned
stability properties are found by constructing a phase portrait of the
ODEs. It is explicitly shown that only unstable equilibrium solutions
are possible with respect to perturbations of the amplitude
eigenvalues. This result serves to suggest further study of the
dynamics (time-evolution) of geon constructs.

The evolution in time of electromagnetic geon solutions is studied in
Sec.~\ref{DynamicStability}.  The evolution of the electromagnetic
geon is studied instead of the gravitational geon because of the
relative ease in computation for the former. With sufficiently
high-frequency electromagnetic waves, the results are expected to
apply equally well to the gravitational case. The method used is
standard perturbation theory modified to accommodate time-averaged
fields. The evolution is achieved by applying a small amplitude
time-dependent perturbation to an equilibrium solution and
simultaneously solving for the time-dependence of the background
metric functions. The problem of time-averaging the background metric
functions can only be done in a meaningful way if it is assumed that
the characteristic frequency of the perturbations vary on a time-scale
much longer than that of the waves comprising the electromagnetic
geon. This is in accordance with the requirement that the background
metric be quasi-stable. However, the results of the analysis show that
the perturbations must vary on the same time-scale as the constituent
waves. This is a contradiction to the original assumption. Hence an
internal inconsistency exists when applying perturbation theory to the
geon problem. In Sec.~\ref{Ch10}, the possible interpretations and
ramifications of this result are discussed.  Since not all of the
requirements for existence of a geon are met, it is not possible to
conclude that an electromagnetic geon or a gravitational geon is a
viable construct. The conclusions are presented in Sec.~\ref{Ch11}.

\setcounter{footnote}{0}
\renewcommand{\thefootnote}{\alph{footnote}}

%
%
\section{Phase Space Analysis}
\label{Ch9}

\noindent
In the high-frequency approximation, the gravitational and
electromagnetic geon problem for a static background metric (on
time-average) reduces to the same set of ordinary differential
equations (ODEs) and boundary conditions given
by\cite{geonletter,geonstanford,geonpaper,AndersonBrill}
\begin{gather} \label{geonwave}
\phi'' + j\, k\, \phi = 0  , \\
 k' = -\phi^2   , \label{geonk} \\
j' = 3 - k^{-2}\left( 1 + \phi^{\prime \,2} \right) ,
\label{geonj} \\
\intertext{and}
\label{ch8_BC}
\begin{split}
\phi(x) \rightarrow 0\,, k(x)\rightarrow 1 \text{ and }
j(x)\rightarrow -\infty \text{ for } x \rightarrow -\infty ,
\displaybreak[0] \\
\phi(x) \rightarrow 0\,, 0 < k(x) < 1 \text{ and }
j(x)\rightarrow -\infty \text{ for } x \rightarrow \infty ,
\end{split}
\end{gather}
where $x$ is a radial coordinate and a prime denotes differentiation
with respect to $x$. Therefore any properties of
Eqs.~\eqref{geonwave}--\eqref{geonj} apply equally well to both the
gravitational and electromagnetic geon case. Any solutions to
\eqref{geonwave}--\eqref{geonj} are necessarily equilibrium solutions
since the background metric is assumed static. In
Sec.~II.\ref{NumericalIntegration}, the numerical solutions presented
in Ref.~\citen{Wheeler55} will be re-examined. The results suggest
further investigation of the numerical solutions is required in order
to determine if the boundary conditions are satisfied. In
Ref.~\citen{AndersonBrill}, it was shown that admissible equilibrium
solutions to Eqs.~\eqref{geonwave}--\eqref{ch8_BC} exist. However, the
stability of these equilibrium solutions were not studied. The
investigation presented in
Sec.~II.\ref{ExistenceAndStabilityOfEquilibriumStates} constructs a
phase portrait of the ODEs from which both the analytic form at
spatial infinity and the stability with respect to perturbations of
the amplitude eigenvalues of any solutions will be obtained.
Knowledge of solution stability provides a basis for investigating the
evolution in time of these solutions.  Unlike other
investigations\cite{Wheeler55,Brill}, we apply a small amplitude
time-dependent perturbation to an equilibrium solution of
\eqref{geonwave}--\eqref{geonj} for the case of an electromagnetic
geon. This is done in Sec.~\ref{DynamicStability}. Solving the
time-dependent perturbation equations leads to a contradiction with
one of the initial assumptions.  The contradiction suggests that
neither an electromagnetic geon nor a gravitational geon is a viable
construct since not all of the requirements for existence of a geon
are met.  This interpretation of the results obtained from this
investigation will be discussed in Sec.~\ref{Ch10}.

\newpage
\subsection{Numerical integration}\label{NumericalIntegration}

\noindent
Wheeler\cite{Wheeler55} originally solved the system
\eqref{geonwave}--\eqref{geonj} by numerical methods in 1955. Since
then computer algorithms have evolved considerably for solving
differential equations.  It is therefore worthwhile to utilize modern
techniques\cite{fn1}
in re-examining those solutions. Even with the
algorithm employed in Ref.~\citen{Wheeler55} for solving the
equations, Wheeler's results are remarkably accurate.

The geon problem (both electromagnetic and gravitational) is reduced
to finding a solution to the autonomous system
\eqref{geonwave}--\eqref{geonj}.  Admissible solutions to
Eqs.~\eqref{geonwave}--\eqref{geonj} are defined as those $\phi(x),\,
j(x)$ and $k(x)$ that satisfy the following criteria:
\begin{enumerate}
\item{\em For large negative $x$}: The wave function $\phi(x)
\rightarrow 0$ and metric function $k(x)~\rightarrow~1$. Under these
conditions $j'(x) \rightarrow 2$.  If $\phi(x),\, j(x)$ and $k(x)$ are
solutions to the autonomous system \eqref{geonwave}--\eqref{geonj},
then so are $\phi(x+a),\, j(x+a)$ and $k(x+a)$ where $a$ is a
constant.  Choosing the integration constant for $j(x)$ to be zero
fixes $a$ and consequently $j(x) \rightarrow 2 x$. This removes any
ambiguity in the start of the integration process. Thus for large
negative $x$, $\phi(x)$ satisfies the equation
\begin{equation} \label{eq9.4}
\frac{\dd^2\phi}{\dd x^2} = 2 x \phi   .
\end{equation}
The approximate solution to \eqref{eq9.4} as given in
Ref.~\citen{Wheeler55} is 
\begin{equation}
\phi(x) = \frac{A}{3} \left(-2x\right)^{-1/4} 
\exp\left(-(-2x)^{3/2}\right)      .
\end{equation}
\item{\em For large positive $x$}: It is required that $\phi(x)
\rightarrow 0$, $0 < k(x) < 1$ and $j(x)$ approach large negative
values. 
\end{enumerate}

The only free parameter is the amplitude $A$ of the wave and this must
be chosen so that the solution fits the boundary conditions. The
nonlinearity of the problem makes it necessary to integrate the
system of equations numerically. The integration is started at
$x=-4$. The initial conditions are as follows:
\begin{align} 
\phi(-4) &= \phi_0   ,\label{eq4.40} \\
\left. \frac{\dd\phi}{\dd x} \right|_{x=-4} &= \left(\frac{1}{16} +
\sqrt{8}\right) \phi_0            ,\\
k(-4) &= 1     ,\\
j(-4) &= -8       . \label{eq4.43}
\end{align}
where $\phi_0$ is to be chosen to give an admissible solution. Those
$\phi_0$ values which yield admissible solutions will be referred to as
eigenvalues of the system \eqref{geonwave}--\eqref{geonj} with initial
conditions \eqref{eq4.40}--\eqref{eq4.43}.

\begin{figure}[p]
%
%
\begin{picture}(360,256)(-30,-20)
\put(0,0){\makebox(360,252)[t]{\includegraphics{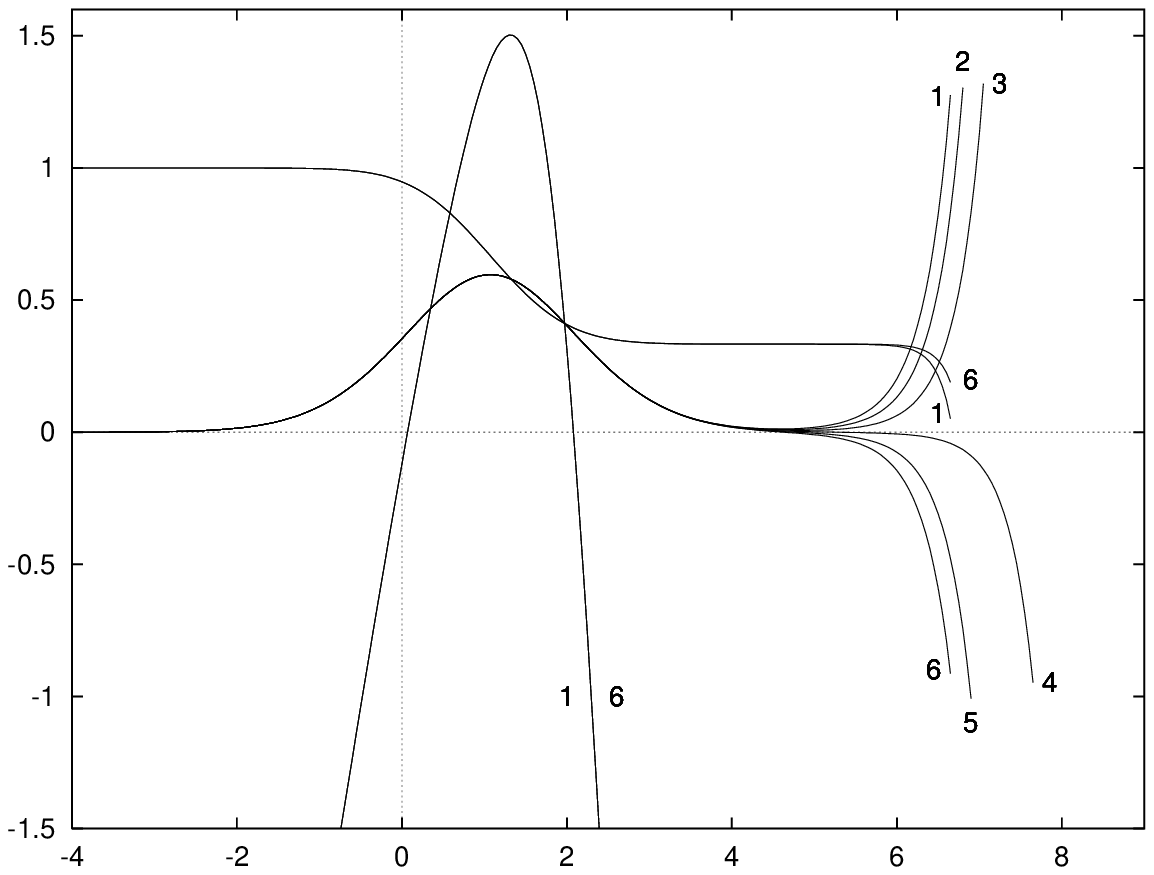}}}
\put(193,-10){\makebox(0,0){$x$}}
\put(60,135){\makebox(0,0){$\phi(x)$}}
\put(100,45){\makebox(0,0){$j(x)$}}
\put(60,210){\makebox(0,0){$k(x)$}}
\end{picture}
%
%
%
\fcaption{Results of the numerical integration for the geon differential
equations~(\ref{geonwave})--(\ref{geonj}). The values of $\phi_0$ for
solution sets 1--6 are summarized in table~\ref{table2}. The
integration started at $x = -4$ and could not proceed beyond $x\simeq
7$ for the given initial values.  The active region begins at $x\simeq
0.12$ and ends at $x\simeq 2.13$. A possible admissible solution lies
between sets 3 and 4. Note that for $j(x)$, curves 1--6 are
indistinguishable.}
\label{fig3}
\end{figure}

The behaviour for $\phi(x)$ as $x\rightarrow \infty$ depends upon the
value of $\phi_0$ (which translates into an initial choice of the
amplitude $A$). Figure~\ref{fig3} illustrates numerically integrated
solutions for values of $\phi_0$ given in table~\ref{table2}. Solution
set 1 shows that for sufficiently large values of $\phi_0$, $\phi(x)$
reaches a positive minimum and then increases exponentially. As
$\phi_0$ is allowed to decrease, the exponential growth of $\phi(x)$
is delayed. This is depicted by solution sets 2 and 3. A further
reduction in $\phi_0$ results in $\phi(x) \rightarrow -\infty$
exponentially as shown in solution sets 4--6. A possible admissible
solution lies between solution sets 3 and 4.  

\begin{table}[p]
\tcaption{Values of $\phi_0$ for solution sets 1--6.}
\centerline{\footnotesize \smalllineskip
\begin{tabular}{cc}\hline\hline 
Solution set & $\phi_0$\\ 
\hline
1 & $\qquad 9.7910\times 10^{-5}$ \\
2 & $\qquad 9.7908\times 10^{-5}$ \\
3 & $\qquad 9.7906\times 10^{-5}$ \\
4 & $\qquad 9.7904\times 10^{-5}$ \\
5 & $\qquad 9.7902\times 10^{-5}$ \\
6 & $\qquad 9.7900\times 10^{-5}$ \\
\hline\hline 
\end{tabular}}
\label{table2}
\end{table}

The mass of the geon inside radius $\rho$ is related to the function
$k(x)$ in the following way:
\begin{equation} M(\rho(x)) = \frac{1}{b} \lambda_0(x) =
\frac{1}{2b} \left(1-k^2\right)       ,
\end{equation} 
with $b=1/Q_1(\infty) = k(\infty)$. This implies that 
\begin{equation} 
0 < k^2(x) \le 1  \mbox{ as } x \rightarrow \infty 
\end{equation} 
in order to have a positive total mass.
The mass factor $k(x)$ gives a positive mass throughout the integrable
region and appears to have a $k(\infty)$ value of approximately
1/3. 
The `active region' can be identified in the $x$ coordinate system as
the region where the function $j(x)$ is positive. In this region, the
function $\phi(x)$ has oscillatory behaviour.  The function $j(x)$ is
positive only for a limited range in the neighbourhood of $x=1$, thus
identifying the active region. In Fig.~\ref{fig3}, the active region
begins at $x\simeq 0.12$ and ends at $x\simeq 2.13$. The first
admissible solution (characterized by $\phi(x)$ having one local
maxima and no local minima) appears to lie between those values of
$\phi_0$ in the range $9.7904\times 10^{-5} < \phi_0 < 9.7906\times
10^{-5}$. Qualitatively, these results are similar to those in
Ref.~\citen{Wheeler55}. The only main difference between the
calculation of Ref.~\citen{Wheeler55} and the present one is that in
Ref.~\citen{Wheeler55}, the first admissible solution appears to lie
in the range $1.03000\times 10^{-4}< \phi_0 <1.03125\times 10^{-4}$
and the active region starts at $x\simeq 4.05$ and ends at $x\simeq
6.02$.

\subsection{Existence and stability of equilibrium states}
\label{ExistenceAndStabilityOfEquilibriumStates} 

\noindent
We are interested in determining the analytical behaviour of the
solutions shown in Fig.~\ref{fig3} as $x\rightarrow\infty$. By
constructing the phase portrait for the differential equations, it
will be possible to determine both the existence and stability
properties of potential admissible solutions. We start by rewriting
equations \eqref{geonwave}--\eqref{geonj} as the set of first order
equations
\begin{align}
u' &= -j\,k\,\phi  ,\label{uprime} \displaybreak[0] \\
\phi' &= u  ,\label{phiprime} \displaybreak[0] \\
 k' &= -\phi^2  ,  \label{kprime} \displaybreak[0] \\
j' &= 3 - k^{-2} + u^2 k^{-2} . \label{jprime}
\end{align}
It is sufficient to look for solutions with the properties
\begin{equation} \label{phasespaceconditions}
\begin{split}
\left.
\begin{array}{l}
 \phi,\,u\rightarrow 0 \\
 k\rightarrow 1
\end{array}
\right\} & \qquad\text{as $x\rightarrow -\infty$,}  \displaybreak[0]
\\ 
\left.
\begin{array}{l}
 \phi,\,u\rightarrow 0 \\
 k\rightarrow \text{constant} > 0
\end{array}
\right\} & \qquad \text{as $x\rightarrow +\infty$} 
\end{split}
\end{equation}
and $j$ remains finite for finite $x$.  In a phase space, the critical
points (or equilibrium points) are characterized by those points where
the derivatives of $u,\,\phi,\,k$ and $j$ are zero. The analytic
behaviour of the solution about a critical point is determined by
analyzing the corresponding linear system in a neighbourhood of that
critical point\cite{JordanSmith}.

The first step is to obtain the critical points of the system
\eqref{uprime}--\eqref{jprime}. One can easily verify that
\begin{align}
   u &=0  ,  \displaybreak[0] \\
\phi &=0  ,  \displaybreak[0] \\
   k &=\pm \frac{1}{\sqrt{3}}  \label{kpm}
\end{align}
is sufficient to satisfy $u'=\phi'=k'=j'=0$.  Thus the coordinates of
the critical point in the phase space are
\begin{align}
   u &=0  ,  \displaybreak[0] \\
\phi &=0  ,  \displaybreak[0] \\
   k &= \frac{1}{\sqrt{3}}  ,   \displaybreak[0] \\
   j &= s,  \qquad s \in \mathbb{R},
\end{align}
where $s$ is any value of $j(x)$. The function $k(x)$ cannot pass
through zero since Eq.~\eqref{geonj} becomes singular. The
positive root of Eq.~(\ref{kpm}) is chosen to ensure positive
$k(x)$, since initially $k(-\infty)=1$.  It is useful to shift the
critical point to the origin using the following transformation:
\begin{equation} \label{phi-f1-transformation}
\begin{matrix}
\phi(x)=f_1(x),\qquad  
u(x)=f_2(x), \\[5pt]  
k(x)=f_3(x) + \dfrac{1}{\sqrt{3}},\qquad
j(x)=f_4(x) + s .
\end{matrix}
\end{equation}
Therefore the field equations become
\begin{align}
f_1' &= f_2  ,  \displaybreak[0] \\
f_2' &= - \left(f_4 + s \right)\left(f_3 + \frac{1}{\sqrt{3}}
\right) f_1   ,  \displaybreak[0] \\
f_3' &= - f_1^2  ,  \displaybreak[0] \\
f_4' &= 3 - \left(1+f_2^2\right) \left(f_3 +
\frac{1}{\sqrt{3}}\right)^{-2}  ,  
\end{align}
with the critical point at $f_1=f_2=f_3=f_4=0$. To linearize the field
equations about this critical point, a MacLaurin series of
$f_i'=f_i'\left(f_1,f_2,f_3,f_4\right),\, i=1,\dots,4$ is taken
to first order and evaluated at the critical point (denoted c.p.\
below), i.e.
\begin{equation}
f_i' =  f_i'\Bigr|_{\text{c.p.}} 
+ \left. \frac{\partial f_i'}{\partial
f_1}\right|_{\text{c.p.}}\hspace{-11pt} f_1
+ \left. \frac{\partial f_i'}{\partial
f_2}\right|_{\text{c.p.}}\hspace{-11pt}  f_2 
+ \left. \frac{\partial f_i'}{\partial
f_3}\right|_{\text{c.p.}}\hspace{-11pt}  f_3 
+ \left. \frac{\partial f_i'}{\partial
f_4}\right|_{\text{c.p.}}\hspace{-11pt}  f_4 + \dotsb .
\end{equation}
Written in matrix form, the linearized field equations are
\begin{equation}
\frac{\dd \mathbf{w}}{\dd x} = \mbox{\sf M}\,\mathbf{w}  ,
\end{equation}
where
\begin{equation}
\mathbf{w} = \begin{pmatrix} f_1 \\ f_2 \\ f_3 \\ f_4 \end{pmatrix} 
\quad \text{ and }\quad  \mbox{\sf M} = \begin{pmatrix}  
                             0 &  1 & 0 	 & 0 \\
	           -s/\sqrt{3} &  0 & 0 	 & 0 \\
			     0 &  0 & 0 	 & 0 \\
   			     0 &  0 & 6\sqrt{3} & 0 
			  \end{pmatrix}  .
\end{equation}
The general solution to the above matrix differential equation is
the eigenvector
\begin{equation}
\mathbf{w} = c_1 \begin{pmatrix} 0 \\ 0 \\ 0 \\ 1 \end{pmatrix}
+ c_2 \left[  \begin{pmatrix} 0 \\ 0 \\ 0 \\ 1 \end{pmatrix} x
+  \begin{pmatrix} 0 \\ 0 \\ \sqrt{3}/18 \\ 0 \end{pmatrix}
      \right] 
       + c_3 \begin{pmatrix} 
                    1 \\ \alpha \\ 0 \\ 0
		    \end{pmatrix} \e^{\alpha\, x}
+ c_4 \begin{pmatrix} 
            1 \\ -\alpha \\ 0 \\ 0
      \end{pmatrix} \e^{-\alpha\, x} \,,
\end{equation}
where $\alpha \equiv \sqrt{-s/\sqrt{3}}$ and $c_i,\, i=1,\dots,4$
are constants. Therefore the solution to the linear system is
\begin{align}
f_1(x) &= c_3 \,\e^{\alpha\, x} + c_4 \,\e^{-\alpha\, x}  ,
\label{PhiLinearSolution} \displaybreak[0] \\
f_2(x) &=    c_3\, \alpha \,\e^{\alpha\, x} 
        -  c_4 \,\alpha \,\e^{-\alpha\, x}  ,\label{uLinearSolution}
 \displaybreak[0] \\  
f_3(x) &= \frac{\sqrt{3}}{18}\, c_2  ,
\label{kLinearSolution}  \displaybreak[0] \\ 
f_4(x) &= c_1 + c_2\, x   . \label{jLinearSolution}
\end{align}
The eigenvector $\mathbf{w}$ shows that in a neighbourhood of the
critical point, the nonlinear system decouples into the disjoint
subspaces $(f_1,\, f_2)$ and $(f_3,\, f_4)$. 

\begin{figure}[p]
%
%
\begin{picture}(360,256)(-20,0)
\put(20,0){\makebox(360,256)[t]{\includegraphics{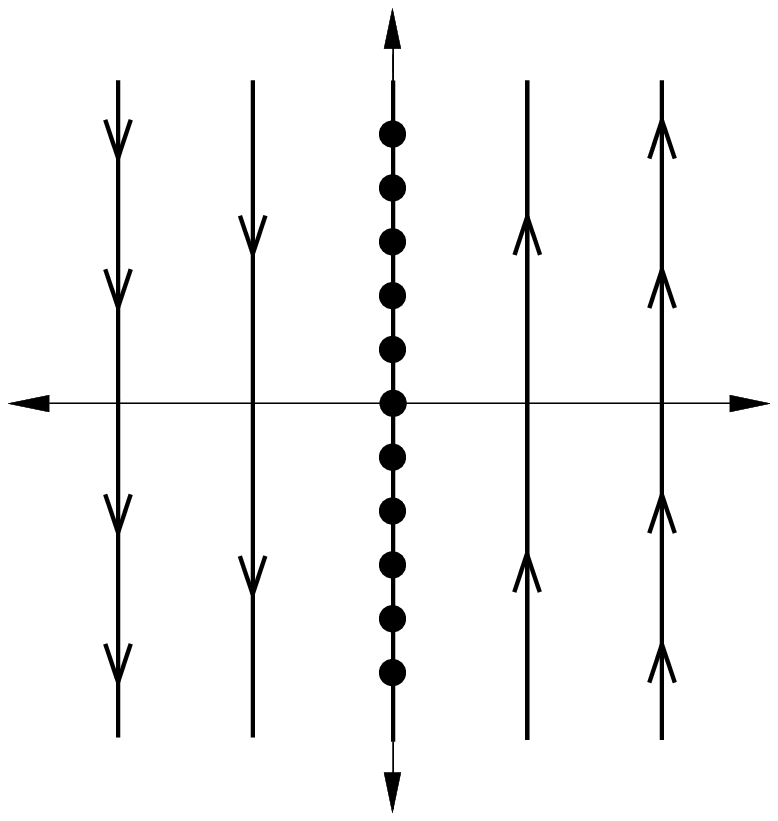}}}
\put(185,235){\makebox(0,0){$f_4$}}
\put(300,120){\makebox(0,0){$f_3$}}
\end{picture}
%
\fcaption{The $(f_3,\, f_4)$ phase space projection illustrates a
non-isolated critical point. The critical point under investigation is
located at the origin. The $f_4$-axis is a continuous set of critical
points in a neighbourhood of the origin. The functions $f_4 \propto j$
and $f_3 \propto k$ behave as a linear function of $x$ and a constant
function respectively in a neighbourhood of the critical point not on
the $f_4$-axis.}
\label{fig4}
\end{figure}

The phase space projection of $(f_3,\, f_4)$ (which is proportional to
$(k,\, j)$) is illustrated in Fig.~\ref{fig4}.
Since $s$ can take any value of $j$, the critical point lies at an
arbitrary position $s \leq \text{max}(j)$ on the curve
$k(x)=\frac{\sqrt{3}}{18}\, c_2 +\frac{1}{\sqrt{3}}$ which is
transformed back to the origin in the $(f_3,f_4)$ subspace. Equations
\eqref{kLinearSolution} and \eqref{jLinearSolution} show that in a
neighbourhood of the critical point, but not on the $f_4$-axis, $f_3$
and $f_4$ behave as a constant and a linear function of $x$
respectively. If one is on the $f_4$-axis in a neighbourhood of the
critical point, then $c_2=0$. Hence $f_4=c_1$ defines a continuous set
of critical points. These critical points are examples of non-isolated
critical points.  The $(f_3,\, f_4)$ projection is insufficient for
determining the existence and stability of admissible solutions since
it only gives information about the functions $j$ and $k$.

\begin{figure}[p]
%
%
\begin{picture}(360,256)(-20,0)
\put(20,0){\makebox(360,256)[t]{\includegraphics{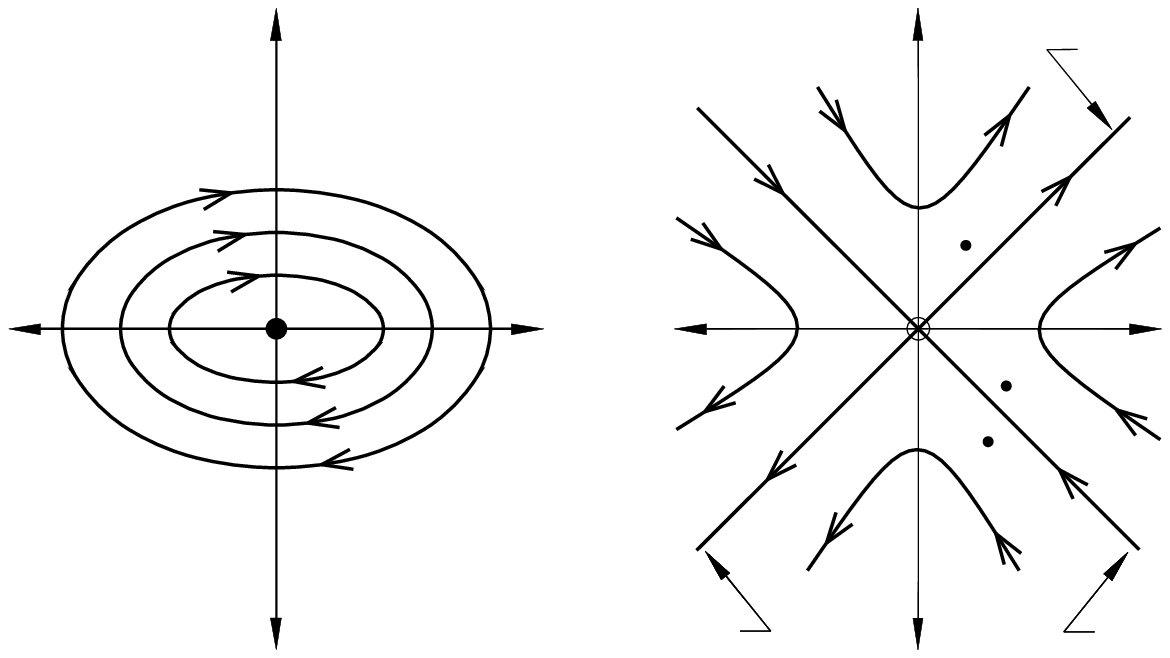}}}
\put(104.5,12){\makebox(0,0){(a)}}
\put(290,12){\makebox(0,0){(b)}}
\put(295,176){\makebox(0,0){1}}
\put(322,135){\makebox(0,0){2}}
\put(300,119){\makebox(0,0){3}}
\put(350,61){\makebox(0,0){$\xi^-$}}
\put(345,229){\makebox(0,0){$\xi^+$}}
\put(223,62){\makebox(0,0){$-\xi^+$}}
\put(90,235){\makebox(0,0){$u$}}
\put(274,235){\makebox(0,0){$u$}}
\put(65,40){\makebox(0,0){$s>0$}}
\put(245,40){\makebox(0,0){$s<0$}}
\put(180,136){\makebox(0,0){$\phi$}}
\put(361,136){\makebox(0,0){$\phi$}}
\end{picture}
%
\fcaption{(a) Illustration of a stable critical point. This type of
critical point occurs when the value of the parameter $s > 0$. (b)
Illustration of an unstable critical point. This type of
critical point occurs when the value of the parameter $s < 0$.}
\label{fig5}
\end{figure}

The nature of the $(f_1,\, f_2)$ subspace\cite{fn2}
(or the $(\phi,\, u)$ subspace) depends upon the parameter $s\in
j$. It will determine the stability properties with respect to
perturbations of the amplitude eigenvalues of any admissible
solutions. However, a 3-dimensional phase space projection in the
coordinates $(\phi,\, u,\, j)$ is necessary to determine the existence
of admissible solutions.  An illustration of the two possible phase
space projections of $(\phi,\, u)$ are shown in Fig.~\ref{fig5}.
Examining \eqref{PhiLinearSolution} and \eqref{uLinearSolution}, if
$s>0$, then $\phi$ and $u$ behave as sinusoidal functions of $x$. This
type of critical point is described as a center (Fig.~\ref{fig5}(a))
and is a {\em stable} critical point. If $s<0$, then $\phi$ and $u$
have an exponential behaviour and the critical point is {\em
unstable}. This type of critical point is described as a saddle point
(Fig.~\ref{fig5}(b)). By following the integration procedure in the
parameter $x$ for solution sets 1 and 6 of Fig.~\ref{fig3}, the
behaviour of the complete nonlinear system can be described.

\begin{figure}[p]
%
%
{\scalebox{0.9}{
\begin{picture}(360,432)(-40,0)
\put(30,110){\makebox(360,432)[t]{\includegraphics{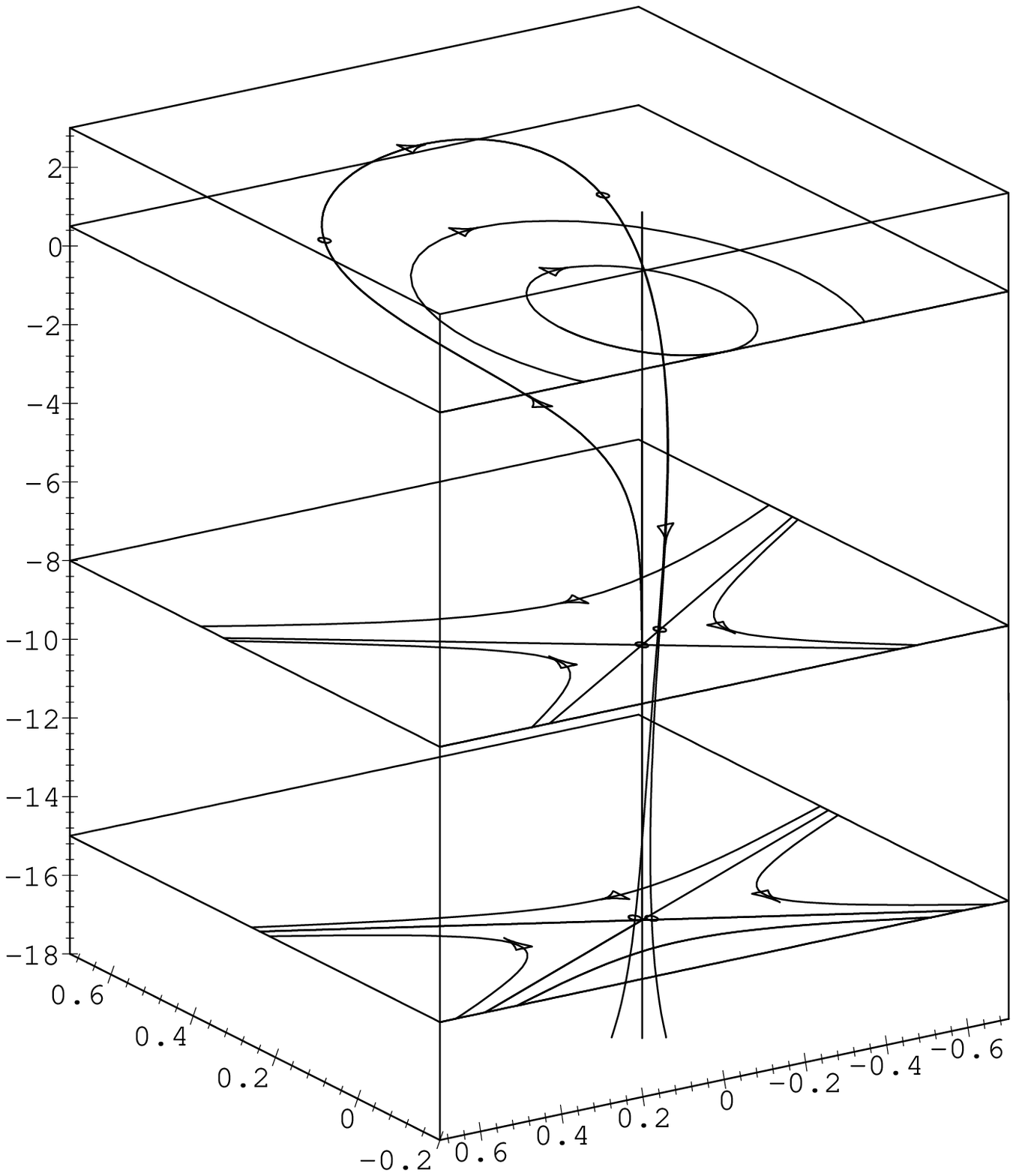}}}
\put(295,-5){\Large $u$}
\put(0,350){\Large $j$}
\put(95,-5){\Large $\phi$}
\put(50,354){A}
\put(50,223){B}
\put(50,112){C}
\put(228,40){\sf 1}
\put(266,40){\sf 6}
\put(200,88){\begin{picture}(20,30)(0,0) 
             \put(0,30){\vector(2,-3){20}}
             \put(0,30){\line(1,0){10}}
             \put(12,27){$\xi^+$}
             \end{picture}}
\put(200,198){\begin{picture}(20,30)(0,0) 
             \put(0,30){\vector(2,-3){20}}
             \put(0,30){\line(1,0){10}}
             \put(12,27){$\xi^+$}
             \end{picture}}
\put(294,126){\begin{picture}(20,30)(0,0) 
             \put(0,30){\vector(2,-3){20}}
             \put(0,30){\line(1,0){10}}
             \put(12,27){$\xi^-$}
             \end{picture}}
\put(280,240){\begin{picture}(20,30)(0,0) 
             \put(0,30){\vector(2,-3){20}}
             \put(0,30){\line(1,0){10}}
             \put(12,27){$\xi^-$}
             \end{picture}}
\end{picture}
}} 
\vspace{36pt}
\fcaption{The 3-dimensional $(\phi,\, u,\,j)$ projection of the
phase space curves for the numerical solution sets 1 and 6 shown in
Fig.~\ref{fig3}. The numerical integration starts in plane B,
proceeds up through plane A and continues down through plane C. The
nature of the critical points along the $j$-axis (origin of the
$(\phi,\,u)$ planes) change from centers ($j>0$) to saddle points
($j<0$) demonstrating the system instability.}
\label{fig6}
\end{figure}

\begin{figure}[p]
\begin{center}
%
%
\begin{picture}(360,256)(15,-10)
\put(15,0){\makebox(360,252)[t]{\includegraphics{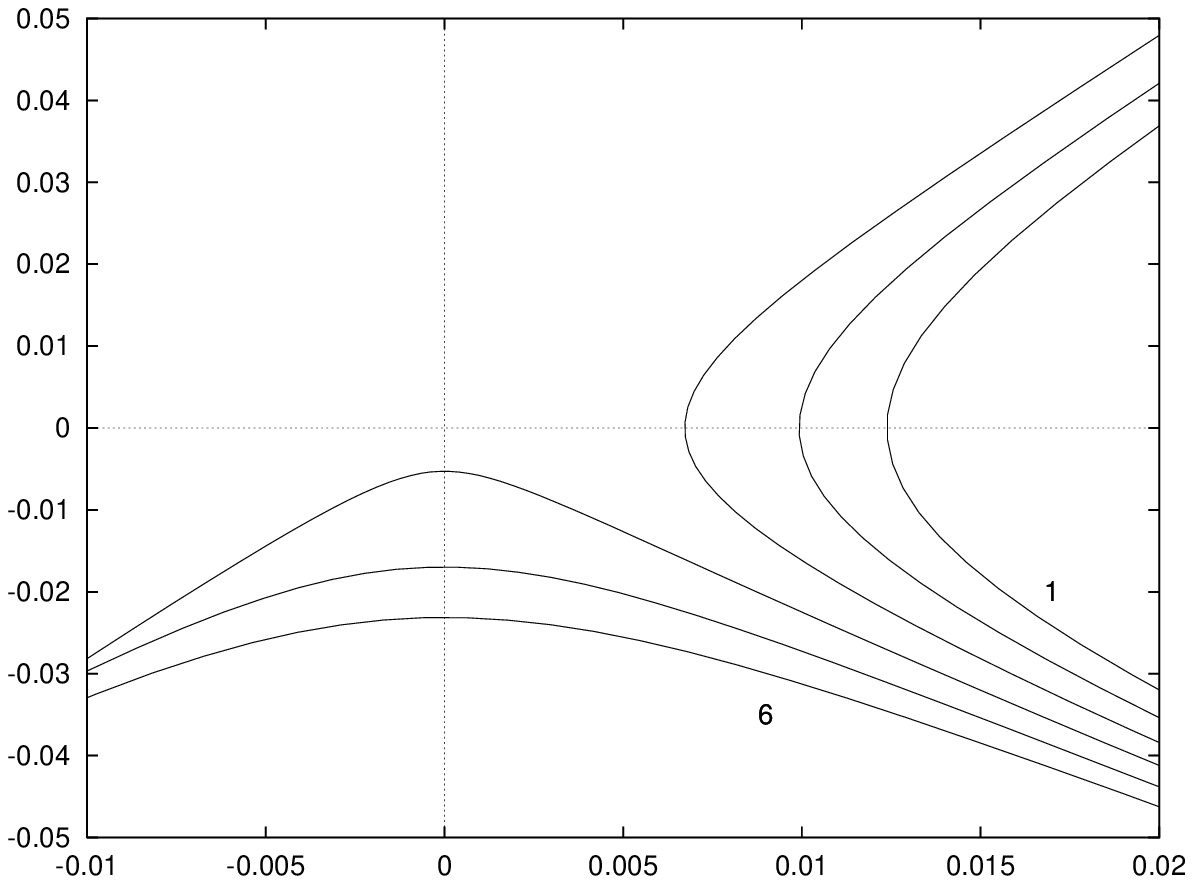}}}
\put(200,-10){\makebox(0,0){\large$\phi$}}
\put(21,130){\makebox(0,0){\Large $u$}}
\end{picture}
\vspace{12pt}
\fcaption{The $(\phi,\,u)$ phase space projection of solution sets
1--6 for $j<-15$.  It has the characteristics of
Fig.~\ref{fig5}(b). This indicates that the critical point at the
origin is unstable as $j\rightarrow -\infty$.}
\label{fig7}
\end{center}
\end{figure}

Figure~\ref{fig6} shows the 3-dimensional $(\phi,\,u,\,j)$ phase space
projection of solution sets 1 and 6. The integration procedure starts
in plane B of Fig.~\ref{fig6} at $j=-8$. In addition, the solution
trajectories start somewhere along the line $u=(1/16+\sqrt{8})\phi_0$
which must necessarily lie to the left of the unstable asymptote
$\xi^+$. One such point is labelled ``1'' in Fig.~\ref{fig5}(b). 
In order for the system to satisfy the boundary
conditions~\eqref{phasespaceconditions}, it is necessary for at least
one solution to flow along the unstable asymptote $\xi^-$ (in the
$j=s\rightarrow -\infty$ plane). If Fig.~\ref{fig5}(b) were a complete
description of the phase space, then it would be impossible for a
solution starting at position ``1'' to cross $\xi^+$. This is a
consequence of the well-known property of autonomous systems that
phase space trajectories do not cross. However, as the integration
process in $x$ continues, the value of $j$ increases from a negative
value to a positive value. Therefore the nature of the critical point
in the 2-dimensional $(\phi,\,u)$ phase planes changes temporarily
from a saddle point to that of a center. Plane A of Fig.~\ref{fig6} is
an illustration of one such critical point. Soon afterward, $j$
decreases to negative values and the critical points are saddle points
once again. However, the solution trajectories have crossed $\xi^+$
and now follow the flow along the unstable asymptote $\xi^-$.  Upon
the transition of the critical point from centers to saddle points,
the asymptotes have been reestablished with solution sets 1 and 6 on
opposite sides of the $\xi^-$ asymptote. The positions where solution
sets 1 and 6 cut the $(\phi,\,u)$ planes for $j<0$ are schematically
illustrated as points ``2'' and ``3'' respectively in
Fig.~\ref{fig5}(b). In Fig.~\ref{fig6}, these positions are most
clearly seen in plane C.  Since trajectories for autonomous systems do
not cross and the trajectories depend continuously on the initial
data, there must be a value of $\phi_0$ for which the trajectory
approaches $\xi^-$ as $j=s\rightarrow -\infty$. The existence of this
trajectory shows that it is possible to find an eigenvalue of $\phi_0$
which satisfies the boundary conditions
\eqref{phasespaceconditions}. However, the nature of the critical
point as $j=s\rightarrow -\infty$ ($x\rightarrow \infty$) is a saddle
point and therefore this eigenvalue solution is an unstable
solution. Since solution set 1 cuts plane C at position ``2'' of
Fig.~\ref{fig5}(b), the flow of the integration process requires this
set to approach $\xi^+$. Similarly, solution set 6 must approach
$\xi^-$, since it cuts plane C at position ``3.\@'' Hence, any small
perturbation of the eigenvalue of $\phi_0$ implies the constant
$c_3\neq 0$ in \eqref{PhiLinearSolution} and
\eqref{uLinearSolution}.  Thus the non-eigenvalue solutions do not
satisfy the boundary conditions. Figure~\ref{fig7} shows the
$(\phi,\,u)$ subspace for the six solution sets of
Fig.~\ref{fig3}. The projection is for $j<-15$. A comparison of
Fig.~\ref{fig7} to Fig.~\ref{fig5}(b) confirms that the admissible
solution is unstable.

The existence of admissible solutions and instability of the
electromagnetic geon system were discussed in Ref.~\citen{Wheeler55}.
However, it was based on the numerical solution curves similar to
Fig.~\ref{fig3}. Performing a phase portrait analysis, we have
formally shown that an eigenvalue does exist\cite{fn3}
which satisfies conditions
\eqref{phasespaceconditions}. We have also shown that this solution
must necessarily be {\em unstable} with respect to perturbations of
the eigenvalue of $\phi_0$. This result suggests that geon constructs
are dynamically unstable, i.e.\ the ensemble of waves must collapse or
explode.  In Ref.~\citen{Wheeler55}, it was also
suggested that a spherical geon would most likely collapse to a
toroidal geon\cite{Ernstgeon}, presumably thought to be more
stable. It was argued by Ernst\cite{Ernstgeon} that the construction
of a toroidal geon could be realized if one had complete knowledge of
a linear geon (which approximates a small segment of the toroidal
geon). Numerical evidence for amplitude eigenvalues in analogy with
the spherical geon were presented in Ref.~\citen{Ernstgeon}.  This
work was extended in Ref.~\citen{perryphd} by performing a phase
portrait analysis. It was found that only unstable admissible
solutions exist, as in the case of the spherical geon.
It has been suggested\cite{AndersonBrill} that this is sufficient for
proclaiming the existence of both types of geons (electromagnetic and
gravitational). However, in Sec.~\ref{introduction}, it was noted that
a true geon must have the property that the ensemble of waves
comprising the geon be confined on a time-scale much longer than the
typical period of the constituent waves. Otherwise it would not be
possible to attribute a structural form to the geon. What is yet to be
determined analytically is the dynamic behaviour of geon constructs
when perturbations are present.
The next section presents a time-dependent perturbation analysis of
the equations that describe the electromagnetic geon in an attempt
to determine this time-scale.

\section{Time-Evolution Analysis of the Electromagnetic Geon}
 \label{DynamicStability}

\noindent
The electromagnetic geon model employed in Ref.~\citen{Wheeler55}
assumed a background metric independent of time. This precludes the
possibility of studying the time-evolution of the individual modes
coupled to a time-evolving background metric. Instead of introducing a
time-dependence into the system of equations and solving the coupled
wave--background system, an alternate method was employed in
Ref.~\citen{Wheeler55} for determining the time-scale of collapse or
`lifetime' of the electromagnetic geon.  The approach was to study the
behaviour of a single photon (borrowing particle physics terminology)
in an effective potential created by the ensemble of photons
comprising the geon. It was expected that photons would leak out of
the potential well in an irreversible dissipation of energy. The model
for the rate of photon leakage was based on the quantum mechanical
process of radioactive alpha-decay. From this model, the probability
of barrier penetration, $\alpha$, (called the `attrition' in
Ref.~\citen{Wheeler55}) has a value of the order
\begin{equation}
\alpha \sim \e^{-2P} ,
\end{equation}
where $P$ is the barrier penetration integral (c.f.\ the Gammow factor
of alpha-decay\cite{krane}). The function $P$ is roughly proportional
to the height of the barrier as encountered by the escaping particle.
A small attrition would translate to a small rate of radiation leakage
out of the effective potential. The time to collapse or the lifetime
of the geon in this model is inversely proportional to the product of
the attrition and angular frequency of the vibrational mode under
consideration. It was estimated using data from the numerical
integration performed in Ref.~\citen{Wheeler55} that the attrition $\alpha
\sim \e^{-1.52 \sqrt{l(l+1)}}$. As the angular momentum $l$ is
increased, the attrition decreases exponentially which suppresses the
photon leakage. Thus it was concluded that it was possible to
construct a geon of virtually infinite lifetime. 

The only similarity between the concept of a geon leaking photons and
radioactive alpha-decay is the shape of the potential well found in
each problem. However, there are important differences to consider.
It is to be stressed that a geon is a classical object built from
classical (massless) fields. Radioactive alpha-decay is a Coulomb
repulsion effect involving massive, electrically charged particles.
The mechanism for the emission of an alpha particle from a nucleus is
the quantum mechanical effect of `tunnelling' through a potential
barrier. In a classical system, it is not possible for a particle to
tunnel though a potential barrier.  In addition, the concept of photon
leakage implies the background metric (or potential barrier) evolves
on some time-scale. An alpha-decay model cannot describe such an
evolution.  The quantum mechanical nature of alpha-decay brings into
question the validity of using such a model for determining the
lifetime of a classical object such as a geon.  There are known
phenomena which represent classical wave penetration of a potential
barrier. For example, the optical phenomenon of frustrated total
internal reflection\cite{eisberg} is such a process. It is important
to note that in both alpha-decay and frustrated total internal
reflection, the potential is supplied by some material and not the
waves themselves, as is necessary for the case of geons. Whether the
analogy exists between these examples of barrier penetration and the
time-evolution of a geon based upon the coupled Einstein--Maxwell
equations is the subject of this section.

Another approach towards determining the lifetime of an
electromagnetic geon is that of Brill\cite{Brill}. The method is to
study the evolution of the ensemble of photons which produce the
effective potential using a thin-shell model for the electromagnetic
geon. It was found that the radial position of the thin shell
underwent a displacement towards collapse. It was also stated that the
rate of collapse was `slow.'

The junction condition problems associated with analyzing thin-shell
geon models has previously been discussed in some detail in
Ref.~\citen{geonpaper}.  In addition, the evolution of the thin-shell
model in Ref.~\citen{Brill} does not allow for leakage of radiation
during the collapse nor does it allow an evolving shell thickness
(evolving active region) as one might expect. The effect of correcting
these deficiencies on the rate of collapse is not clear. A full
understanding of the evolution of a geon must take into account the
evolution of the typical individual modes of vibration coupled to the
evolution of the collective ensemble of waves in a singularity-free
model.

It is evident that models for studying the evolution of geons must be
based on solving the Einstein (or Einstein--Maxwell) field
equations. Intuitive models are not sufficient for describing the true
physical system.  To avoid the interpretation problems associated with
the alpha-decay and thin-shell models and correct for the deficiencies
of each model, the derivation of the electromagnetic geon
equations\cite{Wheeler55} will be modified to permit the study of the
time-evolution of the electromagnetic geon. The electromagnetic geon
will be studied instead of the gravitational geon because of the
relative ease in computation for the former. Equilibrium solutions for
the gravitational and electromagnetic geon are governed by the same
set of ODE's. Therefore it is not expected that the modified
gravitational geon equations in the high-frequency approximation would
yield a significantly different result from the electromagnetic case.

We are interested in following the evolution of the electromagnetic
geon in the radial direction. Observing the time-evolution of the
metric reflects the evolution of the ensemble of electromagnetic waves
comprising the geon. However, it is not sufficient to simply perturb
the background metric functions. It is the electromagnetic waves which
are the source for the gravitational field, hence both the
gravitational and electromagnetic quantities must be perturbed. This
will be done by applying an amplitude perturbation on the
electromagnetic waves comprising the geon in such a way as to induce
the background metric to evolve in time. Frequency perturbations are
not explicitly considered in the following derivation for two main
reasons. Firstly, the stability analysis of the previous section
indicates that the instability of the admissible equilibrium solution
originates from changes in the amplitude eigenvalue (initial
condition). Secondly, it can be shown that a perturbation of the form
$\Omega \rightarrow \Omega + \delta\Omega$ is a special case of the
amplitude perturbation studied below. Restricting
study to the radial direction maintains the field equations in their
simplest form. It should be emphasized that the time-space average of
the electromagnetic disturbance must be incorporated into the
background metric equations to maintain spherical symmetry but still
allow for the solution to evolve in time. This time-averaging problem
will be addressed when the perturbation is applied.  Before the
perturbation analysis is performed, the angle-averaged time-dependent
electromagnetic geon field equations must first be derived. This part
of the derivation follows closely that of Ref.~\citen{Wheeler55}.

The equations presented below are derived in greater detail in
appendix~\ref{App7}. Only an outline of the derivation is given here.
We start by defining the electromagnetic vector potential for one mode
of the electromagnetic waves
\begin{equation} \label{main_vector_potential}
A_\mu = \left(0,\,0,\,0,\,A_\varphi\right) ,
\end{equation}
where
\begin{equation}
A_\varphi = a(r,t) B(\theta)\,, \qquad B(\theta) = \sin\theta
\frac{\dd}{\dd\theta} P_l(\cos\theta)  .
\end{equation}
The function $a(r,t)$ has been left unspecified at this stage.
The time-dependent background metric is
\begin{equation} \label{emgmetric}
\dd s^2 =g_{\alpha\beta}\,\dd x^\alpha \dd x^\beta =  -\e^\nu \dd
t^2 + \e^\lambda \dd r^2 + r^2 \dd \theta^2 + r^2 \sin^2\!\theta \,
\dd\varphi^2  ,
\end{equation}
where
\begin{equation*}
\nu=\nu(r,t),\qquad \lambda=\lambda(r,t) .
\end{equation*}
In the absence of charges and currents, Maxwell's equations in a
curved space-time are
\begin{gather}\label{main_maxwell}
\frac{1}{\sqrt{-g}}\di{\ }{x^\alpha}
\left(\sqrt{-g}F^{\beta\alpha}\right) = 0 , \displaybreak[0] \\[10pt]
F_{\alpha\beta , \gamma} + F_{ \gamma\alpha ,\beta} + 
 F_{\beta \gamma ,\alpha}  = 0  ,
\end{gather}
where $g$ is the determinant of the metric \eqref{emgmetric} and the
Maxwell tensor, $F_{\alpha\beta}$ is related to the four-vector
potential as $F_{\alpha\beta}=A_{\beta , \alpha} - A_{\alpha , \beta}$.
The Einstein equations for the electromagnetic geon are
\begin{equation} \label{main_emgefe}
\G{\mu}{\nu} = 8\pi \left\langle \T{\mu}{\nu} 
\right\rangle  ,
\end{equation}
where $\left\langle \;\boldsymbol{\cdot}\; \right\rangle$ denotes a
time-space average over all $N$ active modes\cite{fn4}
of the electromagnetic waves.  Substituting
\eqref{main_vector_potential} into \eqref{main_maxwell} and
\eqref{main_emgefe}, taking the angle average and finally transforming
to the $\rho=\Omega\, r$ coordinate system yields the wave equation
\begin{equation}\label{wave_time_dependent}
\Omega^2 \ddi{a}{\rho^*} - \Omega^2 l^{*2}\rho^{-2}\left(1 -
\frac{2L}{\rho}\right) Q^2 \, a - \left(1 -
\frac{2L}{\rho}\right)^2 Q^2 \ddi{a}{t^*} = 0
\end{equation}
and the background field equations
\begin{multline} \label{Lp_time_dependent}
\di{L}{\rho^*} = \frac{\kappa_l^2}{2}\left( Q^{-1}\left( \Omega^2
\left\langle \left(\di{a}{\rho^*} \right)^2 
\right\rangle_{\mbox{\tiny T}} + \left\langle \left(\di{a}{t}
\right)^2  \right\rangle_{\mbox{\tiny T}} \right) \right. 
\\ 
+ \Biggl. \Omega^2 
l^{*2}\rho^{-2}\left(1 - \frac{2L}{\rho}\right) Q \Bigl\langle a^2 
\Bigr\rangle_{\mbox{\tiny T}}  \Biggr)  ,
\end{multline}
\begin{equation}\label{Qp_time_dependent}
\di{Q}{\rho^*} = \frac{\kappa_l^2}{\rho - 2L} \left( \Omega^2
\left\langle \left(\di{a}{\rho^*} \right)^2 
\right\rangle_{\mbox{\tiny T}} + \left\langle \left(\di{a}{\rho^*}
\right)^2  \right\rangle_{\mbox{\tiny T}} \right)  ,
\end{equation}
\begin{equation} \label{Lt_time_dependent}
\di{L}{t} = \kappa_l^2 \Omega^2 Q^{-1} \left\langle
\di{a}{\rho^*} \di{a}{t} \right\rangle_{\mbox{\tiny T}}  ,
\end{equation}
where 
\begin{gather}
\di{\ }{\rho^*} \equiv \left(1 - \frac{2L}{\rho}\right)Q  \di{\
}{\rho}  , \displaybreak[0] \\
\ddi{\ }{t^*} = \left(1 - \frac{2L}{\rho}\right)^{-1} Q^{-1} 
\di{\ }{t} \left(\left(1 - \frac{2L}{\rho}\right)^{-1} Q^{-1} 
\di{\ }{t} \right) , \\
\intertext{and}
\kappa_l \equiv \sqrt{\frac{N l^{*2}}{2l+1}}   \qquad l^*\equiv
\sqrt{l(l+1)} \,.
\end{gather}
In the above equations $L(\rho)$ and $Q(\rho)$ are metric functions
(see Eqs.~\eqref{transform_lambda}, \eqref{transform_lambda+nu}) and
the symbol $\left\langle \;\boldsymbol{\cdot}\;
\right\rangle_{\mbox{\tiny T}}$ denotes a time-average is to be
taken. Equations
\eqref{wave_time_dependent}--\eqref{Lt_time_dependent} are the
starting point for developing the dynamic perturbation
(time-evolution) equations.  The $\partial Q/\partial t$ equation
(found from the $\G{\theta}{\theta}=8\pi \left\langle
\T{\theta}{\theta} \right\rangle$ equation) is not used in the
subsequent analysis, but is given in appendix~\ref{App7} for
completeness.

The time-averaged equilibrium solution in Ref.~\citen{Wheeler55} has
the form
\begin{align} \label{equilib_soln_wave}
\kappa_l \,\Omega \, a(\rho,\,t) &= f_0(\rho) \sin \Omega\, t
,\displaybreak[0]\\ 
L(\rho,\, t) &= L_0(\rho)  ,\displaybreak[0]\\
Q(\rho,\, t) &= Q_0(\rho)  . \label{equilib_soln_Q0}
\end{align}
where $f_0(\rho),\, L_0(\rho)$ and $Q_0(\rho)$ are known
functions.\cite{fn5}
We will designate this as the `unperturbed
solution\@.'  Two general forms for the radial perturbation of the
wave function $a(\rho,\, t)$ will be considered. The first is given by
\begin{equation} \label{perturbation1}
\kappa_l \,\Omega \,a(\rho,\, t) = f_0(\rho) \sin \Omega \,t
+ \delta u_1(\rho,\, t) + \delta^2 u_2(\rho,\, t)
+\text{O}\left(\delta^3\right) , \qquad \delta \ll 1 .   
\end{equation}
where $\delta$ is the expansion parameter. (Note that the addition of
a phase constant to $\sin\Omega\,t$ does not affect the results which
follow. Thus the phase constant is set to zero.) As a result, a small
time-dependent perturbation is introduced in the metric functions
\begin{align}
L(\rho,\, t) &= L_0(\rho)  + \delta L_1(\rho,\, t) + \delta^2
L_2(\rho,\, t) +\text{O}\left(\delta^3\right)  ,\displaybreak[0]\\ 
Q(\rho,\, t) &= Q_0(\rho)  + \delta Q_1(\rho,\, t) + \delta^2
Q_2(\rho,\, t) +\text{O}\left(\delta^3\right)
 . \label{perturbationQ} 
\end{align}
The perturbation expansion will be carried out to the first order in
$\delta$. Before the perturbed system is solved in a self-consistent
manner, the problem of time-averaging the functions on the right hand
side of Eqs.~\eqref{Lp_time_dependent}--\eqref{Lt_time_dependent} must
be addressed. From the definition of a time-average and
Eq.~\eqref{perturbation1},
\begin{equation*} \label{time_average_def1}
\begin{split}
\kappa_l^2 \,\Omega^2 \,\bigl\langle a^2(\rho,\,
t)\bigr\rangle_{\mbox{\tiny T}} &\equiv \frac{1}{T} \int_0^T
\kappa_l^2 \Omega^2 a^2(\rho,\, t) \,\dd t 
\displaybreak[0] \\
    &= \frac{1}{T} \int_0^T \Bigl(f^2_0(\rho) \sin^2\!\Omega\, t +
    2 \delta \,u_1(\rho,\, t) f_0(\rho) \sin \Omega\, t \Bigr.
\displaybreak[0] \\
    & \qquad\qquad + \Bigl.\delta^2 \left( u^2_1(\rho,\,t) +
    2 u_2(\rho,\, t) f_0(\rho) \sin \Omega\, t  \right)+
    \mbox{O}\left(\delta^3\right)\Bigr)   
    \,\dd t  
\displaybreak[0] \\
    &=  \frac{1}{2} f^2_0(\rho) +  \frac{1}{T} \int_0^T \Bigl(2 \delta
       u_1(\rho,\,t) f_0(\rho) \sin \Omega\, t \Bigr. 
\displaybreak[0] \\
    & \qquad\qquad + \Bigl.\delta^2 \left( u^2_1(\rho,\,t) + 2
       u_2(\rho,\, t) f_0(\rho) \sin \Omega\, t \right) +
       \mbox{O}\left(\delta^3\right)\Bigr) \,\dd t 
        . 
\end{split}
\end{equation*}
where $T=2\pi\Omega^{-1}$ is the period of the electromagnetic
waves. In order to develop the perturbation analysis, it is
necessary to make some assumptions about the function
$u_1(\rho,t)$. The presence of unevaluated integrals on the right hand
side of the differential equations
\eqref{Lp_time_dependent}--\eqref{Lt_time_dependent} would not make it
possible to proceed with the analysis beyond this point. Let us
suppose the time-dependence of $u_1(\rho,t)$ was sinusoidal and its
characteristic frequency was of the order $\Omega$ of the
electromagnetic waves. In this case, the time-dependence is lost to
all orders in $\delta$ upon time-averaging. In essence, this
assumption on $u_1(\rho,\,t)$ precludes the possibility of a
time-dependent evolution of the system. This is not satisfactory. To
maintain a time-dependence after time-averaging, another time-scale
will be introduced into the problem. Suppose the time-dependence of
$u_1(\rho,t)$ was again sinusoidal and its characteristic frequency
was of the order $\omega \ll\Omega$. In this case, $u_1(\rho,t)$ is
approximately constant over the short time period $T=2\pi\Omega^{-1}$
of the electromagnetic waves and thus is constant in the time-average
integral. Evaluating the time-average of $a^2(\rho,\,t)$ under this
assumption yields
\begin{align} \label{time_average_def2}
\kappa_l^2 \,\Omega^2 \,\bigl\langle a^2(\rho,\,
t)\bigr\rangle_{\mbox{\tiny T}} 
    &= \frac{1}{T} \int_0^T \Bigl(f^2_0(\rho) \sin^2\!\Omega\, t +
    2 \delta \,u_1(\rho,\, t) f_0(\rho) \sin \Omega\, t
\Bigr. \displaybreak[0]\notag\\ 
    & \qquad\qquad + \Bigl.\delta^2 \left( u^2_1(\rho,\,t) +
    2 u_2(\rho,\, t) f_0(\rho) \sin \Omega\, t  \right)\Bigr)  
    \,\dd t + \mbox{O}\left(\delta^3\right)
\displaybreak[0]\notag\\
    &=  \frac{1}{2} f^2_0(\rho) + 2\delta u_1(\rho,\,t) f_0(\rho)
       \frac{1}{T} \int_0^T  \sin \Omega\, t \,\dd t 
\displaybreak[0]\notag\\
    &\phantom{=} + \delta^2 u^2_1(\rho,\,t) \frac{1}{T} \int_0^T \dd t
       + 2 u_2(\rho,\, t) f_0(\rho) \int_0^T  \sin \Omega\, t \,\dd t
       + \mbox{O}\left(\delta^3\right) 
\displaybreak[0] \notag\\
    &=  \frac{1}{2} f^2_0(\rho) + \delta^2 u^2_1(\rho,\,t)
    + \mbox{O}\left(\delta^3\right)  . 
\end{align}
Therefore the time-dependence is not present until the second order in
$\delta$. This is sufficient to proceed with the time-evolution of the
system, since each order in the expansion parameter $\delta$ must be
set to zero. To {\em first} order in $\delta$
\begin{align}
 &\kappa_l^2 \,\Omega^2 \,\bigl\langle a^2(\rho,\,
 t)\bigr\rangle_{\mbox{\tiny T}} = \frac{1}{2}  f_0^2 
+ \mbox{O}\left(\delta^2\right)  . \label{avg_a*a} 
\displaybreak[0]\\   
\intertext{Similarly, the remaining time-averages are}
 &\kappa_l^2 \,\Omega^2
\,\left\langle \left(\di{a}{\rho^*} \right)^2
\right\rangle_{\mbox{\tiny T}} = \frac{1}{2} \left( \diff{f_0}{\rho^*}
\right)^2  
+ \mbox{O}\left(\delta^2\right)  ,
\displaybreak[0]\\
 &\kappa_l^2 \,\Omega^2 \,\left\langle
\left(\di{a}{t} \right)^2 \right\rangle_{\mbox{\tiny T}} =
\frac{1}{2}\Omega^2 f_0^2  
+\mbox{O}\left(\delta^2\right)   ,
\displaybreak[0]\\
 &\kappa_l^2 \,\Omega^2 \,\left\langle
\di{a}{t} \di{a}{\rho^*} \right\rangle_{\mbox{\tiny T}} =
 \mbox{O} \left(\delta^2 \right)
 . \label{avg_atap} 
\end{align}
Substitution of \eqref{perturbation1}--\eqref{avg_atap} into
\eqref{wave_time_dependent}--\eqref{Lt_time_dependent}, 
expanding to first order in $\delta$ and setting each order in
$\delta$ to zero yields the unperturbed equations
\eqref{f0eqn}--\eqref{Q0eqn}.
The properties of the unperturbed
equations
\begin{gather} \label{f0eqn}
\ddiff{f_0}{\rho^*} + \Biggl( 1 - l^{*2} Q_0^2\rho^{-2}
\left(1-\frac{2L_0}{\rho}\right) \Biggr) f_0 = 0 , 
\displaybreak[0]\\
\diff{L_0}{\rho^*} = \frac{1}{2Q_0} \left( f_0^2 +
\left(\diff{f_0}{\rho^*}\right)^2  + l^{*2} Q_0^2\rho^{-2}
\left(1-\frac{2L_0}{\rho}\right) f_0^2 \right) , \label{L0eqn}
\displaybreak[0]\\
\diff{Q_0}{\rho^*} = \left(\rho - 2L_0\right)^{-1}\left( f_0^2 +
\left(\diff{f_0}{\rho^*}\right)^2 \right)  , \label{Q0eqn}
\end{gather}
are known from Ref.~\citen{Wheeler55} and therefore can be
used in the analysis of the first order equations.
Setting the first order part of the wave equation
\eqref{wave_time_dependent} to zero yields
\begin{equation} \label{first_wave}
A(\rho,\,t)  \sin \Omega \,t +
B(\rho,\, t) \cos \Omega \,t + C(\rho,\, t) = 0
 , 
\end{equation}
where
\begin{flalign}  
 \mbox{} &A(\rho,\,t)  \equiv \Omega \Biggl( \Biggr. 
\left( Q_1(\rho,\,t) 
\left(1 -\frac{2L_0}{\rho} \right)
-2\rho^{-1} L_1(\rho,\,t) Q_0 \right) 
\times
\displaybreak[0] \notag \\
 &  \times \biggl( \biggr. 2\rho^{-2}  Q_0 \diff{f_0}{\rho}
\left( L_0 -\rho\diff{L_0}{\rho} \right)  + Q_0  \ddiff{f_0}{\rho} 
\left(1 -\frac{2L_0}{\rho} \right)
 + \diff{f_0}{\rho}  \diff{Q_0}{\rho}  
\left(1 -\frac{2L_0}{\rho} \right) 
\biggl. \biggr)
\displaybreak[0] \notag \\
 & \quad + 2\rho^{-3} l^{*2} Q_0 \left( Q_0  L_1(\rho,\,t) 
-  \rho  Q_1(\rho,\,t) \left(1 -\frac{2L_0}{\rho} \right)
\right) f_0
+ Q_0 
\left(1 -\frac{2L_0}{\rho} \right)
\times
\displaybreak[0] \notag \\
 & \quad  \times \Biggl( \Biggr.  2\rho^{-2}  \diff{f_0}{\rho}\left(
 Q_1  \left( L_0  -\rho\diff{L_0}{\rho} \right) + Q_0 \left(
 L_1(\rho,\,t)  -\rho\di{\ }{\rho}L_1(\rho,\,t) \right) \right)
\displaybreak[0] \notag \\
 & \quad +  \left( 
\left(1 -\frac{2L_0}{\rho} \right)
\di{\ }{\rho}Q_1(\rho,\,t) - 2\rho^{-1}
L_1(\rho,\,t)\diff{Q_0}{\rho}\right)\diff{f_0}{\rho}  
\displaybreak[0] \notag \\
 & \quad +  \left( Q_1(\rho,\,t) 
\left(1 -\frac{2L_0}{\rho} \right)
-2\rho^{-1} L_1(\rho,\,t) Q_0 \right)\ddiff{f_0}{\rho} \Biggl. \Biggr)
\displaybreak[0] \notag \\
 & \quad + \frac{f_0}{\rho - 2 L_0} \left( 2 L_1(\rho,\,t) - \rho
 \left(1 -\frac{2L_0}{\rho} \right) Q_0^{-1} Q_1(\rho,\,t) \right)
\displaybreak[0] \notag \\
 & \quad + \left(1 -\frac{2L_0}{\rho} \right)^{-1} Q_0^{-1} \left(
 Q_1(\rho,\,t)  
\left(1 -\frac{2L_0}{\rho} \right)
-2\rho^{-1} L_1(\rho,\,t) Q_0 \right) f_0
\Biggl. \Biggr) , \label{mathfrakA}
\end{flalign}  
\begin{equation} \label{mathfrakB} 
B(\rho,\, t) \equiv  f_0 \left( 2 \rho^{-1}
\left(1-\frac{2L_0}{\rho}\right)^{-1} \di{\ }{t}L_1(\rho,\,
t)  - Q_0^{-1} \di{\ }{t}Q_1(\rho,\, t) \right) 
\end{equation}
and
\begin{align}  \label{mathfrakC}
C(\rho,\, t) \equiv \ddi{\ }{\rho^*}u_1(\rho,\, t)  -
l^{*2} Q_0^2\rho^{-2} \left(1-\frac{2L_0}{\rho}\right)  u_1(\rho,\, t)
-\Omega^{-2} \ddi{\ }{t}u_1(\rho,\, t)  .
\end{align}

In order to satisfy the first order wave equation \eqref{first_wave},
the conditions $A(\rho,\,t) =0$, $B(\rho,\, t) =0$ and
$C(\rho,\, t) =0$ must be imposed. This is justified since
$\sin\Omega\,t$ and $\cos\Omega\,t$ are independent functions in the
approximation where $A(\rho,\,t)$, $B(\rho,\, t)$ and
$C(\rho,\, t)$ are slowly varying functions of time. We
will focus upon the latter equation, since it is the simplest of the
three equations. Setting \eqref{mathfrakC} to zero yields the equation
\begin{equation} \label{full_u1eqn}
\ddi{\ }{\rho^*}u_1(\rho,\, t)  - l^{*2} Q_0^2\rho^{-2}
\left(1-\frac{2L_0}{\rho}\right)  u_1(\rho,\, t) -\Omega^{-2}
\ddi{\ }{t}u_1(\rho,\, t)= 0 .  
\end{equation}
To solve this equation, we will first use the method of separation of
variables. Later, this restriction will be removed. Let
\begin{equation} \label{u(p)T(t)}
u_1(\rho,\,t) = u(\rho) T(t) .
\end{equation}
Substituting \eqref{u(p)T(t)} into \eqref{full_u1eqn} and dividing by
$u(\rho) T(t)$ yields the two ordinary differential equations
\begin{gather}
\ddiff{u(\rho)}{\rho^*}  - l^{*2} Q_0^2\rho^{-2}
\left(1-\frac{2L_0}{\rho}\right)  u(\rho)  + \beta u(\rho) =0 ,
\label{u(p)_only}
\\
\ddiff{T(t)}{t} +\beta\, \Omega^2 T(t) = 0 ,  
\label{T(t)_only}
\end{gather}
where $-\beta$ is the separation constant. The solution to
\eqref{T(t)_only} is 
\begin{equation}
T(t) = c_3 \sin(\omega t + c_4)\,, \qquad c_3,c_4 \text{ -- constants,}
\end{equation}
where we have chosen $\beta\equiv \omega^2/\Omega^2$, $\omega
\ll \Omega$. Making this choice for $\beta$ satisfies the requirement
of \eqref{time_average_def2} for a meaningful time-average. 

In order to solve Eq.~\eqref{u(p)_only}, it is necessary to apply
the high angular momentum approximation. It is therefore necessary to
transform ${C}(\rho,\, t)$ to the $x$ coordinate system and
expand in inverse powers of $l^{*1/3}$ as is done for the unperturbed
system\cite{Wheeler55}. The transformation in Ref.~\citen{Wheeler55} for
the unperturbed functions is repeated here for convenience
\begin{align} \label{x_transformation}
x &= \left(\rho^* - l^{*}\right) l^{*-1/3}, \\
\dd\rho^* &\equiv  l^{*1/3} \dd x   ,\displaybreak[0] \notag\\ 
\rho &=  l^{*} + l^{*1/3} r_0(x) + \dotsb   , \displaybreak[0]
\notag\\ 
L_0 &=  l^{*} \lambda_0(x) + l^{*2/3} \lambda_1(x) + l^{*1/3}
\lambda_2(x) + \dotsb   , \displaybreak[0] \label{ell_expansion} \\ 
Q_0 &=  1/k(x) + l^{*-1/3} q_1(x) + l^{*-2/3} q_2(x)
+ \dotsb   , \displaybreak[0] \notag\\
f_0 &=  l^{*1/3} \phi(x) +  \phi_1(x) + l^{*-1/3} \phi_2(x)
+ \dotsb   . \notag \\
\intertext{In addition, the function $u(\rho)$ must be expanded in a
similar manner, i.e.}
u &=  l^{*1/3} \mu_0(x) +  \mu_1(x) + l^{*-1/3} \mu_2(x)
+ \dotsb   . \label{u_expansion} 
\end{align}
After a lengthy computation, the asymptotic expansion of
\eqref{u(p)_only} yields 
\begin{equation} \label{one_third}
l^{*1/3} \left( \dfrac{\omega^2}{\Omega^2} -
\dfrac{1-2\lambda_0(x)}{k^2(x)} \right) \mu_0(x) + \text{O}(1)  = 0
 . 
\end{equation}
In order for \eqref{one_third} to be satisfied for large
arbitrary $l^{*}$ (in the limit $l^{*}\rightarrow \infty$), each order
of $l^{*1/3}$ must be set to zero. Since setting $\mu_0(x)=0$ implies
the absence of electromagnetic wave perturbations, the bracketed term
must be zero. It is known from the unperturbed system that
\cite{Wheeler55}
\begin{equation} \label{lambda_k}
\lambda_0(x) = \frac{1}{2}\left(1-k^2(x)\right) .
\end{equation}
Substituting \eqref{lambda_k} into \eqref{one_third} and setting the
bracketed term to zero yields the relation
\begin{equation} \label{contradiction}
\omega^2 =  \Omega^2
\end{equation}
which must hold in order to satisfy the condition
$C(\rho,t)=0$. However, Eq.~\eqref{contradiction} is a
{\em contradiction} to the original assumption $\omega \ll
\Omega$. Because of the presence of the contradiction, it is not
necessary to solve the remaining field equations.

We arrived at this contradiction through the assumption that
\eqref{full_u1eqn} could be solved by separating variables
(Eq.~\eqref{u(p)T(t)}). The same result is obtained if $u_1(\rho,\,
t)$ is not separable as is shown below. Modifying \eqref{u_expansion}
as follows
\begin{equation}
u_1 =  l^{*1/3} \mu_0(x,\,t) +  \mu_1(x,\,t) + l^{*-1/3} \mu_2(x,\,t)
+ \dotsb   . \label{new_u_expansion} 
\end{equation}
and expanding \eqref{full_u1eqn} in inverse powers of $l^{*1/3}$
yields
\begin{equation}
-l^{*1/3}\left( \Omega^{-2} \ddi{\ }{t}\mu_0(x,\,t) -
\dfrac{1-2\lambda_0(x)}{k^2(x)} \right)
\mu_0(x,\,t) + \text{O}(1) = 0
\end{equation}
to lowest order in $l^{*1/3}$. Setting each order in $l^{*1/3}$ to
zero requires $\mu_0(x,\,t)$ to satisfy the differential equation
(using \eqref{lambda_k})
\begin{equation}
\ddi{\ }{t}\mu_0(x,\,t) + \Omega^2 \mu_0(x,\,t) =0  .
\end{equation}
The solution is
\begin{equation} \label{mu(x,t)_solution}
\mu_0(x,\,t) = c_5(x) \sin(\Omega\,t + c_6(x)) .
\end{equation}
The characteristic frequency of $\mu_0(x,\,t)$ is $\Omega$
which contradicts the assumption $u_1(\rho,\,t) \sim \mu_0(x,\,t)
\sim \omega \ll \Omega$. 

Before we proceed with the discussion of the perturbation analysis
based on \eqref{perturbation1}--\eqref{perturbationQ},
a second form for the wave function perturbation should be
considered. Consider the perturbation in the form
\begin{align} \label{alt_perturbation}
\kappa_l \,\Omega \,a(\rho,\, t) &= \left( f_0(\rho) + \delta
u_1(\rho,\,t) \right) \sin \Omega \,t +
\mbox{O}\left(\delta^2\right) , 
\end{align}
where the characteristic frequency of $u_1(\rho,\,t)$ is of order
$\omega \ll \Omega$. This form can be interpreted as a slowly evolving
amplitude of the rapidly varying function $\sin \Omega\,t$.
Equations~\eqref{perturbation1} and \eqref{alt_perturbation}, in
addition to the assumptions placed on $u_1(\rho,\,t)$, cover the
entire range of possibilities for these types of perturbations (for
example, a perturbation of the form $a(\rho,\,t) \rightarrow a(\rho +
\delta \xi(\rho,\,t),\,t)$ reduces to \eqref{alt_perturbation}).
Evaluation of the time-averages yields
\begin{align}
 &\kappa_l^2 \,\Omega^2 \,\bigl\langle a^2(\rho,\,
 t)\bigr\rangle_{\mbox{\tiny T}} = \frac{1}{2}
 f_0^2 +  \delta f_0 u_1 + \mbox{O}\left(\delta^2\right)   ,
\label{alt_avg_a*a} 
\\  
 &\kappa_l^2 \,\Omega^2 \,\left\langle \left(\di{a}{\rho^*} \right)^2
 \right\rangle_{\mbox{\tiny T}} = \frac{1}{2} \left( \diff{f_0}{\rho^*}
 \right)^2 +  \delta \diff{f_0}{\rho^*} \di{u_1}{\rho^*} +
 \mbox{O}\left(\delta^2\right)   ,  
\\ 
 &\kappa_l^2 \,\Omega^2 \,\left\langle \left(\di{a}{t} \right)^2
 \right\rangle_{\mbox{\tiny T}} = \frac{1}{2}\Omega^2 f_0^2
 + \delta \, \Omega^2 f_0 u_1 + \mbox{O}\left(\delta^2\right)  , 
\\  
 &\kappa_l^2 \,\Omega^2 \,\left\langle \di{a}{t} \di{a}{\rho^*}
 \right\rangle_{\mbox{\tiny T}} =  \delta \frac{1}{2} 
 \diff{f_0}{\rho^*} \di{u_1}{t} + \mbox{O}
 \left(\delta^2\right)  . \label{alt_avg_atap} 
\end{align} 
Hence, the time-dependence of the time-averaged functions becomes
manifest at the first order in $\delta$. This greatly increases the
mathematical complexity of the system. The detailed analysis for this
system is derived in appendix~\ref{App8} for the case of
$u_1(\rho,\,t)$ separable. From this analysis, the same contradiction
results for the characteristic frequency of $u_1(\rho,\,t)$ as that
found from the analysis based upon Eq.~\eqref{perturbation1}. For
nonseparable $u_1(\rho,\,t)$, the differential equations become
unmanageable. Hence, it was not possible to obtain a conclusive
result. However, due to the similar nature of the systems based upon
Eqs.~\eqref{perturbation1} and \eqref{alt_perturbation}, there is
no reason to suspect a different result for the nonseparable case.

%
%
\section{Discussion}
        \label{Ch10}
\noindent
We have seen from the analysis of the static solution (Sec.~\ref{Ch9})
that the confinement of geon constructs demands an absolutely critical
choice of initial condition (amplitude eigenvalue). The slightest
deviation from that choice leads to a totally unconfined
structure. While some might argue that the confined structure being
indicated with the critical initial condition is already
satisfactory,\cite{AndersonBrill} the general experience with
solitonic structures is one of essentially confined solutions in the
neighbourhood of the best choice of critical
condition.\cite{CoopRosen} The failure to find a family of
near-confinement in the case of the geon already raises suspicions as
to its viability.

In the previous section, the time-evolution of an electromagnetic geon
equilibrium solution was studied with the objective of determining the
time-scale of collapse away from the equilibrium configuration.
Perturbations of the form \eqref{perturbation1} and
\eqref{alt_perturbation} were analyzed under certain assumptions.  The
problem of time-averaging the source terms (right hand side) of the
field equations \eqref{Lp_time_dependent}--\eqref{Lt_time_dependent}
requires the characteristic frequency $\omega$ of the perturbation
term of the wave function $u_1(\rho,\,t)$ to be much less than the
characteristic frequency $\Omega$ of the unperturbed solution.
Without this assumption, the time-dependence of the perturbations is
lost upon time-averaging, to all orders in the expansion.
Hence, the assumption $u_1(\rho,\,t) \sim\omega \sim \Omega$ is not
satisfactory for studying the time-evolution of geons.

The assumption $u_1(\rho,\,t) \sim\omega \ll \Omega$, for both forms
of the perturbation, solves the time-averaging problem in a simple
manner. With this assumption, the differential equations maintain a
time-dependence after the time-average has been taken over the typical
period of the high-frequency waves. The perturbation analysis leads to
the requirement $\omega\sim \Omega$ in order for the field equations
to be satisfied. This is a contradiction to the original assumption
$\omega \ll \Omega$. Since all the possible combinations for the form
of the perturbation (and assumptions placed on $u_1(\rho,\,t)$) have
been explored, the possible interpretations of the results of
Sec.~\ref{DynamicStability} are given below.

A reasonable interpretation of the contradiction in the time-evolution
analysis is that the condition of slow evolution of the background
cannot be satisfied. Since not all of the required conditions are
satisfied, it is not possible to construct a geon comprised of
high-frequency waves. It could be argued that an electromagnetic geon
could be built from low-frequency waves. This has not been ruled out
by our model since it only accommodates high-frequency waves. However,
a gravitational geon necessarily must be constructed from
high-frequency waves, for otherwise the effective stress-energy would not
be of the correct order of magnitude to create the background
gravitational field binding the waves
\cite{geonletter,geonstanford,geonpaper}. 
Therefore the gravitational geon studied in this paper is subject to
the same fate as its high-frequency electromagnetic counterpart.

A geon with a rapidly evolving background metric, where the background
is somehow regarded as being distinct from the small amplitude waves
on the background, is conceptually unsound.  The definition presented
in Sec.~\ref{introduction} for this type of geon requires the
background solution of the Einstein or Einstein--Maxwell field
equations be quasi-stable on a time-scale much longer than the period
of the constituent waves. If the background metric evolves away from
the equilibrium configuration on the same time-scale as the
constituent waves, one cannot speak of the waves binding
gravitationally. Under these circumstances, there is no geon structure
to identify.  Apart from this argument, the realization of a geon with
a rapidly varying background metric $\gamma_{\mu\nu}$ is problematic
for another reason. As discussed in Ref.~\citen{geonpaper}, if a
spherically symmetric background is allowed to vary harmonically with
frequency $\Omega$ comparable to the frequency of the gravitational or
electromagnetic waves, one expects a parametric resonance
\cite{Arnold} for the modes with $\omega_n =n \Omega/2$, with $n \in
\mathbb{N}$. The strength of the resonance is a maximum for $n=1$ and
decreases rapidly as $n$ increases. In the limit of a static
background, the resonance phenomenon disappears. Accordingly, on the
basis of studies of perturbations of black holes and relativistic
stars \cite{Chandrasekhar}, it is expected that in the case of a
stationary axisymmetric background metric describing a rapidly
rotating geon, the resonance phenomenon between the perturbations and
the background metric occurs.  In the general case of a time-dependent
and rapidly varying background metric $\gamma_{\mu\nu}\left( t,
\vec{x}\right)$ without symmetries, it is not known how to decompose
metric perturbations on a complete set playing the role of the tensor
spherical harmonics in the spherical case, or even how to define
frequencies in the strong curvature region. However, if such concepts
can be given a meaning, it seems reasonable to expect some kind of
resonance phenomena between the background metric and its wave
perturbations. All these resonance phenomena certainly do not
contribute to the realization of a stable configuration, but rather
are associated with instabilities that tend to disrupt the system.

The contradiction which arises in the perturbation analysis of
Sec.~\ref{DynamicStability} is interpreted as a breakdown of the
model, i.e.\ the perturbation model is not able to describe the
evolution of the physical system. It might be argued that an
alternative method of implementing the time-evolution (not using
perturbative methods) may be better suited to determine the time-scale
of evolution. For example, it may be possible to develop an exact
numerical solution of the full Einstein (or Einstein--Maxwell)
equations without the splitting of the metric into a background and
waves on the background or taking time-averages. Considering the
complexity involved in analyzing the simple perturbative model of
Sec.~\ref{DynamicStability}, any new model would undoubtedly be more
complex to solve.  However, it should be possible to approximate any
exact method with an appropriate perturbation expansion (as is
presented in Sec.~\ref{DynamicStability}). Therefore, it is
appropriate to consider the contradiction in a physical context, as
was discussed earlier.

At this point, we recall the original motivation which led the authors
to re-open the issue of geons and their viability. One of the
authors\cite{Cooperstock,Cooperstock2,Cooperstock3} had been led to
propose a new hypothesis regarding the localization of energy in
general relativity, namely that energy was localized in regions of
non-vanishing energy-momentum tensor $T^{\mu\nu}$. There were various
factors leading to this. Firstly, the energy-momentum conservation
laws
\begin{equation}\label{tmunu}
           T^{\mu\nu}_{\quad ;\nu}=0 
\end{equation}
are devoid of content in vacuum, producing the empty identity $0=0$.
However, when \eqref{tmunu} is re-expressed as a vanishing ordinary
divergence to create a global form of the conservation law involving
the introduction of pseudotensors, it is used to compute supposed
fluxes of gravitational field energy in vacuum where the originating
law is devoid of content. It was proposed that the ambiguity of the
pseudotensorial flux vectors actually reflects the illegitimate
injection of supposed physical content where none actually
exists. Furthermore, it was shown that for Kerr--Schild metrics, all
components of the gravitational pseudotensors
vanish\cite{GursesGursey} and gravitational plane waves can be
expressed in Kerr--Schild form.  Since a wave is plane in a relatively
small region, this is further support to the belief that waves of
gravity are not actually carriers of energy in vacuum, in accord with
the localization hypothesis.  Other aspects to support the hypothesis
had been outlined including the relationship to the important earlier
papers of Nissani and Leibowitz \cite{Nissani}, the basis for
non-excitation of a Feynman detector and the work of Virbhadra
\cite{Virbhadra} which showed that localization of energy is confined
to the $T^{\mu\nu}$ regions for static and stationary spacetimes. The
gravitational geon remained an outstanding challenge to the
localization hypothesis since a purely gravitational non-singular
construct displaying an unambiguous mass would be a clear
counter-example. The present work adds new support for the hypothesis
apart from the value of understanding this basic construct.  From
another viewpoint, the results which we found in this paper are not
surprising. From studies extending over 65 years, it was recognized
that non-singular soliton structures to model elementary particles are
not easily achieved and they are successful only with a careful
mixture of different types of fields (see Ref.~\citen{CoopRosen} for a
review with earlier references contained therein).  The
electromagnetic geon depends only on the electromagnetic field and its
own gravity while the gravitational geon is even more restrictive,
being a purely gravitational construct. In the light of earlier
studies, it is not surprising that such simple ingredients should
resist compactification.
 
%
%
\section{Conclusions}
        \label{Ch11}
\noindent
The construction of a satisfactory gravitational geon model requires
an asymptotically flat, self-consistent solution of the Einstein field
equations which meets the regularity conditions for a singularity-free
space-time. Furthermore, it must be demonstrated that the evolution in
time of the geon must take place on a time-scale much longer than the
characteristic period of the constituent waves (quasi-stability
property).

To satisfy these conditions, it was proposed
\cite{geonletter,geonstanford,geonpaper} that a satisfactory
gravitational geon model must be constructed in a manner similar to
that of Wheeler's\cite{Wheeler55} electromagnetic geon.  This type of
model for the gravitational geon is in contrast to the thin-shell
model of Brill and Hartle\cite{BrillHartle}. In order to construct a
gravitational geon in principle, it was previously established
\cite{Isaacson,geonletter,geonstanford,geonpaper} that gravitational
waves of high-frequency were necessary.  The application of the
high-frequency approximation reduced the gravitational and
electromagnetic geon problem to the same set of ordinary differential
equations and boundary conditions. Since the background metric is
initially assumed static, any solutions are necessarily equilibrium
solutions. From a phase portrait analysis of the ordinary differential
equations governing gravitational and electromagnetic geons, it was
possible to determine both the existence and stability properties of
equilibrium solutions with respect to purturbations of the amplitude
eigenvalues. It was found that admissible equilibrium solutions were
unstable to changes in the amplitude eigenvalues.  Since a basic
requirement for the existence of both types geon is the
quasi-stability property, an investigation of the time-evolution of an
electromagnetic geon was performed.  In contrast to other
investigations, a small amplitude time-dependent perturbation to an
equilibrium solution was applied.  The time-averaging problem is
solved by assuming the characteristic frequency of the perturbations
vary on a time-scale much longer than that of the waves comprising the
electromagnetic geon. This is in accordance with the requirement that
the background metric be quasi-stable. Solving the time-dependent
perturbation equations leads to the characteristic frequency of the
perturbations being of the same order in magnitude as the waves
comprising the electromagnetic geon. This is a contradiction to the
original assumption. Thus it could not be shown that the
time-evolution of the electromagnetic geon proceeds on a slow
time-scale using standard perturbation theory modified for
time-averaged fields. With not all of the requirements for the
existence of an electromagnetic geon being satisfied, it cannot be
concluded that an electromagnetic or a gravitational geon is a viable
entity.

Given the results as applied to the gravitational geon, such a
construct cannot be considered a counter-example to the energy
localization hypothesis as discussed in
Ref.~\citen{Cooperstock,Cooperstock2,Cooperstock3}.


\nonumsection{Acknowledgements}
\noindent
The authors would like to thank Dr.\ S. Bohun for helpful discussions.
This research was supported, in part, by a grant from the Natural
Sciences and Engineering Research Council of Canada and a Natural
Sciences and Engineering Research Council Postgraduate Scholarship
(GPP).

\appendix{Time-Dependent Electro\-mag\-netic Geon Equa\-tions}
\label{App7} 
%
%
\noindent
We start by defining the electromagnetic vector potential for one
mode of the electromagnetic waves
\begin{equation} \label{vector_potential}
A_\mu = \left(0,\,0,\,0,\,A_\varphi\right) , 
\end{equation}
where
\begin{equation}
A_\varphi = a(r,t) B_l(\theta)\,, \qquad B_l(\theta) = \sin\theta
\frac{\dd}{\dd\theta} P_l(\cos\theta)  .
\end{equation}
The time-dependent metric is
\begin{equation} \label{app7emgmetric}
\dd s^2 =g_{\alpha\beta}\,\dd x^\alpha \dd x^\beta =  -\e^\nu \dd
t^2 + \e^\lambda \dd r^2 + r^2 \dd \theta^2 + r^2 \sin^2\!\theta \,
\dd\varphi^2  ,
\end{equation}
where
\begin{equation*}
\nu=\nu(r,t),\qquad \lambda=\lambda(r,t) .
\end{equation*}
In the absence of charges and currents, Maxwell's equations in a
curved space-time are
\begin{gather}\label{app7_maxwell}
\frac{1}{\sqrt{-g}}\di{\ }{x^\alpha}
\left(\sqrt{-g}F^{\beta\alpha}\right) = 0 , \\[10pt]
F_{\alpha\beta , \gamma} + F_{ \gamma\alpha ,\beta} + 
 F_{\beta \gamma ,\alpha}  = 0  ,
\end{gather}
where $g$ is the determinant of the metric \eqref{app7emgmetric} and
the Maxwell tensor, $F_{\alpha\beta}$ is related to the four-vector
potential as $F_{\alpha\beta}=A_{\beta , \alpha} - A_{\alpha ,
\beta}$.  The only nontrivial equation is for $\alpha=\varphi$ in
\eqref{app7_maxwell}. It yields the wave equation
\begin{equation}
\ddi{a}{r^*} - \frac{l(l+1)}{r^2} \e^\nu a -
\e^{\nu-\lambda}\ddi{a}{t^*} = 0  ,
\end{equation}
where
\begin{align}
\di{\ }{r^*}&=\e^{(\nu-\lambda)/2}\di{\ }{r} , 
       & \ddi{\ }{r^*} &=\e^{(\nu-\lambda)/2}\di{\ }{r}
	  \left(\e^{(\nu-\lambda)/2}\di{\ }{r}\right) , \label{r_star}
\\ 
\di{\ }{t^*}&=\e^{(\nu-\lambda)/2}\di{\ }{t} , 
       & \ddi{\ }{t^*} &=\e^{(\nu-\lambda)/2}\di{\ }{t}
	  \left(\e^{(\nu-\lambda)/2}\di{\ }{t}\right) . \label{t_star}
\end{align}
The Einstein equations for the electromagnetic geon are
\begin{equation} \label{emgefe}
\G{\mu}{\nu} = 8\pi \left\langle \T{\mu}{\nu}
\right\rangle  ,
\end{equation}
where $\left\langle \;\boldsymbol{\cdot}\; \right\rangle$ denotes a
time-space average over all $N$ active modes of the electromagnetic
waves. In the equations below, the energy-momentum tensor for a single
mode of electromagnetic radiation is given by
\begin{equation}
T^{\hspace{14pt} \nu}_{({\mbox{\tiny I}})\;\mu}\equiv
\dfrac{1}{4\pi}\left( F_{\mu\sigma}F^{\sigma\nu} - \dfrac{1}{4}
F_{\alpha\beta}F^{\alpha\beta} \delta_\mu^\nu \right)  ,
\end{equation}
with $F_{\alpha\beta}$ defined above.  We will only evaluate the angle
average of $\T{\mu}{\nu}$.  The time-average will be dealt with in the
main text (Sec.~\ref{DynamicStability}). In addition to the three angle
averages\cite{Wheeler55,fn6}
\begin{equation} \label{angle_avg_tt}   
\Bigl\langle 
\T{t}{t}\Bigr\rangle_{\mbox{\tiny TA}} = \frac{N}{2} \int_0^\pi
\Bigl\langle T^{\hspace{14pt} t}_{({\mbox{\tiny
I}})\;t}\Bigr\rangle_{\mbox{\tiny T}} 
\sin\theta \, \dd\theta           ,
\end{equation} 
\begin{equation} 
\Bigl\langle  \T{r}{r}\Bigr\rangle_{\mbox{\tiny TA}} = \frac{N}{2} 
\int_0^\pi \Bigl\langle T^{\hspace{14pt} r}_{({\mbox{\tiny
I}})\;r}\Bigr\rangle_{\mbox{\tiny T}} 
\sin\theta \, \dd\theta               ,
\end{equation}
\begin{equation} \label{angle_avg_theta_theta}
\Bigl\langle  \T{\theta}{\theta} \Bigr\rangle_{\mbox{\tiny TA}}
=  \Bigl\langle  
\T{\varphi}{\varphi} \Bigr\rangle_{\mbox{\tiny TA}} 
 =  \frac{N}{2} \int_0^\pi
\Bigl\langle T^{\hspace{14pt} \theta}_{({\mbox{\tiny I}})\;\theta} 
+T^{\hspace{14pt}\varphi}_{({\mbox{\tiny
I}})\;\varphi}\Bigr\rangle_{\mbox{\tiny T}} 
\sin\theta \, \dd\theta                   ,
\end{equation}
there is an additional average which represents the radial flow of
energy
\begin{equation} \label{angle_avg_rt}   
\Bigl\langle 
\T{r}{t}\Bigr\rangle_{\mbox{\tiny TA}} = \frac{N}{2} \int_0^\pi
\Bigl\langle 
T^{\hspace{14pt} t}_{({\mbox{\tiny I}})\;r}\Bigr\rangle_{\mbox{\tiny T}}
\sin\theta \, \dd\theta           .
\end{equation} 
Evaluating \eqref{angle_avg_tt}--\eqref{angle_avg_rt} using
\eqref{vector_potential} and the integrals of appendix~\ref{App5} one
obtains 
\begin{align}
\Bigl\langle 
\T{t}{t}\Bigr\rangle_{\mbox{\tiny TA}} &= -\frac{N l(l+1)}{8\pi
r^2(2l+1)} \left(\e^{-\nu} \Bigl\langle
a^2_{,t}\Bigr\rangle_{\mbox{\tiny T}} + \e^{-\lambda}  \Bigl\langle
a^2_{,r}\Bigr\rangle_{\mbox{\tiny T}} + \frac{l(l+1)}{r^2}
\Bigl\langle a^2\Bigr\rangle_{\mbox{\tiny T}} \right) ,
\\
\Bigl\langle 
\T{r}{r}\Bigr\rangle_{\mbox{\tiny TA}} &= \frac{N l(l+1)}{8\pi
r^2(2l+1)} \left(\e^{-\nu} \Bigl\langle
a^2_{,t}\Bigr\rangle_{\mbox{\tiny T}} + \e^{-\lambda}  \Bigl\langle
a^2_{,r}\Bigr\rangle_{\mbox{\tiny T}} - \frac{l(l+1)}{r^2}  \Bigl\langle
a^2\Bigr\rangle_{\mbox{\tiny T}} \right) ,
\\
\Bigl\langle 
\T{\theta}{\theta}\Bigr\rangle_{\mbox{\tiny TA}} &= \frac{N
l^2(l+1)^2}{8\pi r^4 (2l+1)} \Bigl\langle
a^2\Bigr\rangle_{\mbox{\tiny T}} ,
\\
\Bigl\langle 
\T{r}{t}\Bigr\rangle_{\mbox{\tiny TA}} &= -\frac{N
l(l+1)}{4\pi \e^\nu r^2 (2l+1)} \Bigl\langle
a_{,r} a_{,t} \Bigr\rangle_{\mbox{\tiny T}} .
\end{align}
The components of the left hand side of \eqref{emgefe} are 
\begin{align}
\G{t}{t} &=\mbox{} -r^{-2} + r^{-2}\e^{-\lambda} -
r^{-1}\e^{-\lambda}\lambda_{,r} 
\\
\G{r}{r} &=\mbox{} -r^{-2} + r^{-2}\e^{-\lambda} +
r^{-1}\e^{-\lambda}\nu_{,r} 
\\
\G{\theta}{\theta} &=\G{\varphi}{\varphi} = \frac{1}{2}\left(
\e^{-\lambda}\left(r^{-1}\nu_{,r} -r^{-1}\lambda_{,r}  + \nu_{,rr}
-\frac{1}{2} \lambda_{,r}\nu_{,r} + \frac{1}{2} \nu^2_{,r}\right)
\right. \notag \\
& \qquad \qquad\qquad  + \left. \e^{-\nu}\left(
\frac{1}{2} \lambda_{,t}\nu_{,t} -\lambda_{,tt} - \frac{1}{2}
\lambda^2_{,t} \right) \right)
\\
\G{r}{t} &= -r^{-1}\e^{-\nu} \lambda_{,t}  .
\end{align}
The final step in obtaining the time-dependent electro\-mag\-netic
geon field equations is to make the transformation to the $\rho$
coordinate system. In addition to the transformation
\begin{equation} \label{transform_rho}
r = \frac{\rho}{\Omega} ,
\end{equation}
we introduce the two metric functions $L(\rho,\, t)$ and $Q(\rho,\,
t)$ through the defining equations
\begin{gather} 
e^{-\lambda} \equiv 1 - \frac{2L(\rho,\, t)}{\rho}  ,
\label{transform_lambda} \\
e^{\lambda+\nu} \equiv Q^2(\rho,\, t)  , 
\label{transform_lambda+nu} \\  
e^{\nu} = \left(1 - \frac{2L(\rho,\, t)}{\rho}\right)Q^2(\rho,\, t) 
 . \label{transform_nu} 
\end{gather} 
The operator $\di{\ }{r^*}$ has the following form in the $\rho$
coordinate system
\begin{equation}
\di{\ }{r^*}=\e^{(\nu-\lambda)/2}\di{\ }{r} = \Omega \left(1 -
\frac{2L}{\rho}\right)Q  \di{\ }{\rho} .  
\end{equation} 
By defining 
\begin{equation}
\di{\ }{\rho^*} \equiv \left(1 - \frac{2L}{\rho}\right)Q  \di{\
}{\rho}  ,
\end{equation}
the operators of \eqref{r_star} simply transform as
\begin{equation} 
\di{\ }{r^*} = \Omega \di{\ }{\rho^*}\,, \qquad \ddi{\ }{r^*} = \Omega^2
\ddi{\ }{\rho^*} . 
\end{equation}
The operator $\ddi{\ }{t^*}$ of \eqref{t_star} transforms as
\begin{equation} \label{ttrans_star}
\ddi{\ }{t^*} = \left(1 - \frac{2L}{\rho}\right)^{-1} Q^{-1} 
\di{\ }{t} \left(\left(1 - \frac{2L}{\rho}\right)^{-1} Q^{-1} 
\di{\ }{t} \right) .
\end{equation}
After applying the transformation \eqref{transform_rho} and
Eqs.~\eqref{transform_lambda}--\eqref{ttrans_star}, a lengthy but
straightforward computation yields the wave equation
\begin{equation} \label{app7_wave_p_eq}
\Omega^2 \ddi{a}{\rho^*} - \Omega^2 l^{*2}\rho^{-2}\left(1 -
\frac{2L}{\rho}\right) Q^2 \, a - \left(1 -
\frac{2L}{\rho}\right)^2 Q^2 \ddi{a}{t^*} = 0
\end{equation}
and the background field equations
\begin{multline}
\di{L}{\rho^*} = \frac{\kappa_l^2}{2}\left( Q^{-1}\left( \Omega^2
\left\langle \left(\di{a}{\rho^*} \right)^2 
\right\rangle_{\mbox{\tiny T}} + \left\langle \left(\di{a}{t}
\right)^2  \right\rangle_{\mbox{\tiny T}} \right) \right. 
\\ 
+ \Biggl. \Omega^2 
l^{*2}\rho^{-2}\left(1 - \frac{2L}{\rho}\right) Q \Bigl\langle a^2 
\Bigr\rangle_{\mbox{\tiny T}}  \Biggr)  ,  
\end{multline} 
\begin{equation} 
\di{Q}{\rho^*} = \frac{\kappa_l^2}{\rho - 2L} \left( \Omega^2
\left\langle \left(\di{a}{\rho^*} \right)^2 
\right\rangle_{\mbox{\tiny T}} + \left\langle \left(\di{a}{\rho^*}
\right)^2  \right\rangle_{\mbox{\tiny T}} \right)  ,
\end{equation}
\begin{equation} \label{app7_Lt_eq}
\di{L}{t} = \kappa_l^2 \Omega^2 Q^{-1} \left\langle
\di{a}{\rho^*} \di{a}{t} \right\rangle_{\mbox{\tiny T}} 
\end{equation}
and
\begin{multline}
\ddi{L}{t} + 4\rho^{-1}\left( \di{L}{t}\right)^2 - Q^{-1}
\di{L}{t}\di{Q}{t}  = \frac{1}{2} \left(1 - \frac{2L}{\rho}\right)^2
Q^2 \rho  \times \\[3pt] \times \Bigl(  {\cal A}(\rho)+ {\cal B}(\rho)
- 2\kappa^2 l^{*2} \Omega^4 \rho^{-4} \Bigl\langle a^2 
\Bigr\rangle_{\mbox{\tiny T}} \Bigr)  .
\end{multline}
In the above equations we have defined
\begin{equation}
\kappa_l \equiv \sqrt{\frac{N l^{*2}}{2l+1}} \, \qquad l^*\equiv
\sqrt{l(l+1)}  ,
\end{equation}
\begin{align}
{\cal A}(\rho) &\equiv 2 \Omega^2 \rho^{-1}\Biggl(
 Q^{-1} \di{Q}{\rho} 
\left(1 - \frac{2L}{\rho}\right) + 2\rho^{-1}\left(\rho^{-1} L
-\di{L}{\rho} \right)\Biggr)  \\
\intertext{and}
{\cal B}(\rho) &\equiv \Omega^2  \left(1-\frac{2L}{\rho}\right) 
\left( 2 \makebox(0,24){}\left( 2 Q^{-1}\left(\ddi{Q}{\rho} -
Q^{-1}\left(\di{Q}{\rho}\right)^2 
\right) \right.\right. 
\displaybreak[0] \notag \\
 &  \qquad -
2\rho^{-2}\left(1-\frac{2L}{\rho}\right)^{-1} \left(\rho^{-1} L
-\di{L}{\rho} \right) 
\displaybreak[0] \notag \\
&  \qquad + 4\rho^{-2}\left(1-\frac{2L}{\rho}\right)^{-2}
\left(\rho^{-1} L -\di{L}{\rho} \right)^2 
\displaybreak[0] \notag \\ 
&  \qquad  + \left.
2\rho^{-2}\left(1-\frac{2L}{\rho}\right)^{-1} 
\left( \di{L}{\rho} - \rho^{-1} L - \rho\ddi{L}{\rho}\right)\right) 
\displaybreak[0] \notag \\
& \qquad - \left(2\rho^{-1}\left(1-\frac{2L}{\rho}\right)^{-1}
\left( \di{L}{\rho} - \rho^{-1} L \right)\right) \times 
\displaybreak[0] \notag \\
& \qquad \times \left( 2 Q^{-1}\di{Q}{\rho} -
2\rho^{-1}\left(1-\frac{2L}{\rho}\right)^{-1} 
\left( \di{L}{\rho} - \rho^{-1} L \right)\right) 
\displaybreak[0] \notag \\
&  \qquad  + \left. \left( 2 Q^{-1}\di{Q}{\rho} -
2\rho^{-1}\left(1-\frac{2L}{\rho}\right)^{-1} 
\left( \di{L}{\rho} - \rho^{-1} L \right) \right)^2 \right)  .
\end{align}
Equations \eqref{app7_wave_p_eq}--\eqref{app7_Lt_eq} are
Eqs.~\eqref{wave_time_dependent}--\eqref{Lt_time_dependent} of 
Sec.~\ref{DynamicStability}. 


\appendix{Perturbation Analysis of a Slowly Varying Amplitude}
\label{App8}
%
%
\noindent
To investigate the time-evolution of the unperturbed solution, the
equilibrium solution
\eqref{equilib_soln_wave}--\eqref{equilib_soln_Q0} will be perturbed
by allowing the coefficient of $\sin \Omega\, t$ to become a slowly
varying function of time as compared to the period $2\pi\Omega^{-1}$
of the electromagnetic waves. In addition, it will be assumed that
$u_1(\rho,\,t)$ is a separable function (i.e.\
$u_1(\rho,\,t)=u(\rho)T(t)$). Under this assumption, 
\eqref{alt_perturbation} becomes
\begin{equation} 
\kappa_l \,\Omega \,a(\rho,\, t) = \left( f_0(\rho) + \delta
u(\rho)T(t) \right) \sin \Omega \,t ,
\end{equation}
where the characteristic frequency of $u(\rho)T(t)$ is of order
$\omega \ll \Omega$.  This introduces a small time-dependent
perturbation in the metric functions
\begin{align}
L(\rho,\, t) &= L_0(\rho) + \delta L_1(\rho,\, t) , \\
Q(\rho,\, t) &= Q_0(\rho) + \delta Q_1(\rho,\, t)  .
\end{align}
It is stated without proof that for $u_1(\rho,\,t)$ separable, the
field equations impose the relations $u(\rho)=f_0(\rho)$ and
$T(t)=\sin(\omega\,t)$. Therefore the analysis in this appendix is
carried out for the perturbation
\begin{equation} \label{apph_perturbation}
\kappa_l \,\Omega \,a(\rho,\, t) = \left( f_0(\rho) + \delta
f_0(\rho)\sin\omega t\right) \sin \Omega \,t\,, \qquad \omega \ll
\Omega\,, \quad \delta \ll 1 . 
\end{equation}
The perturbation expansion will be taken to the first order in
$\delta$.  We have assumed that the coefficient $f(\rho,\, t) \equiv
f_0(\rho) + \delta f_0(\rho)\sin\omega t$ of $\sin\Omega\, t$ in
\eqref{apph_perturbation} varies on a time-scale much longer than
that of $\sin\Omega\, t$. Therefore $f(\rho,\, t)$ is approximately
constant over the short time period $T=2\pi\Omega^{-1}$ of the
electromagnetic waves. Therefore, evaluation of the time-averages
\eqref{alt_avg_a*a}--\eqref{alt_avg_atap} yields
\begin{align}
 &\kappa_l^2 \,\Omega^2 \,\bigl\langle a^2(\rho,\,
 t)\bigr\rangle_{\mbox{\tiny T}} = \frac{1}{2}
 f_0^2 \left( 1 + 2 \delta \sin \omega t \right) +
 \mbox{O}\left(\delta^2\right)  , \\
 &\kappa_l^2 \,\Omega^2 \,\left\langle \left(\di{a}{\rho^*} \right)^2
 \right\rangle_{\mbox{\tiny T}} = \frac{1}{2}\left( \diff{f_0}{\rho^*}
 \right)^2 \left( 1 + 2 \delta \sin \omega t \right) +
 \mbox{O}\left(\delta^2\right)   , \\ 
 &\kappa_l^2 \,\Omega^2 \,\left\langle \left(\di{a}{t} \right)^2
 \right\rangle_{\mbox{\tiny T}} = \frac{1}{2}\Omega^2 f_0^2
 \left( 1 + 2 \delta \sin \omega t \right) +
 \mbox{O}\left(\delta^2\right)   ,\\ 
 &\kappa_l^2 \,\Omega^2 \,\left\langle \di{a}{t} \di{a}{\rho^*}
 \right\rangle_{\mbox{\tiny T}} = \frac{1}{2}\delta\omega f_0
 \diff{f_0}{\rho^*}  \cos \omega t + \mbox{O}
 \left(\delta^2\right)  . \label{apph_avg_atap}
\end{align}
Substitution of \eqref{apph_perturbation}--\eqref{apph_avg_atap} into
\eqref{wave_time_dependent}--\eqref{Lt_time_dependent}, 
expanding to first order in $\delta$ and setting each order to zero
yields the unperturbed equations \eqref{f0eqn}--\eqref{Q0eqn} and the
first order equations \eqref{apph_first_wave}--\eqref{apph_lteq}. 
Setting the first order part
of the wave equation \eqref{wave_time_dependent} to zero yields
\begin{equation} \label{apph_first_wave}
A(\rho,\,t)  \sin \Omega \,t +
B(\rho,\, t) \cos \Omega \,t = 0  ,
\end{equation}
where
\begin{align}  
\mbox{} & A(\rho,\,t) \equiv
\Omega ^2 \Biggl( \Biggr.
\omega^2\Omega^{-2} f_0  +  \left(1 - l^{*2} \rho^{-2} Q_0^2 
\left(1 -\frac{2L_0}{\rho} \right)
 \right) \,
f_0 +  Q_0 \left(1 -\frac{2L_0}{\rho} \right) \times
\displaybreak[0] \notag \\
 & \qquad \times \biggl( \biggr. 2\rho^{-2}  Q_0 \diff{f_0}{\rho}
\left( L_0 -\rho\diff{L_0}{\rho} \right)  + Q_0  \ddiff{f_0}{\rho} 
\left(1 -\frac{2L_0}{\rho} \right)
\displaybreak[0] \notag \\
 & \qquad + \diff{f_0}{\rho}  \diff{Q_0}{\rho}  
\left(1 -\frac{2L_0}{\rho} \right) 
\biggl. \biggr) \Biggl. \Biggr)
 \sin \omega \,t
 + \Omega^2 \Biggl( \Biggr.  Q_0 
\left(1 -\frac{2L_0}{\rho} \right)
\times
\displaybreak[0] \notag \\
 & \qquad  \times \biggl( \biggr.  2\rho^{-2}  \diff{f_0}{\rho}\left(
 Q_1  \left( L_0  -\rho\diff{L_0}{\rho} \right) + Q_0 \left(
 L_1(\rho,\,t)  -\rho\di{\ }{\rho}L_1(\rho,\,t) \right) \right)
\displaybreak[0] \notag \\
 & \qquad +  \left( 
\left(1 -\frac{2L_0}{\rho} \right)
\di{\ }{\rho}Q_1(\rho,\,t) - 2\rho^{-1}
L_1(\rho,\,t)\diff{Q_0}{\rho}\right)\diff{f_0}{\rho}  
\displaybreak[0] \notag \\
 & \qquad +  \left( Q_1(\rho,\,t) 
\left(1 -\frac{2L_0}{\rho} \right)
-2\rho^{-1} L_1(\rho,\,t) Q_0 \right)\ddiff{f_0}{\rho} \biggl. \biggr)
\displaybreak[0] \notag \\
 & \qquad + 2\rho^{-3} \left( l^{*2} Q_0^2  L_1(\rho,\,t) 
- l^{*2} \rho  Q_0 Q_1(\rho,\,t) \left(1 -\frac{2L_0}{\rho} \right)
\right) f_0
\displaybreak[0] \notag \\
 & \qquad +\left( Q_1(\rho,\,t) 
\left(1 -\frac{2L_0}{\rho} \right)
-2\rho^{-1} L_1(\rho,\,t) Q_0 \right) \biggl( \biggr.
2\rho^{-2}  \diff{f_0}{\rho} 
 Q_0  \left( L_0  -\rho\diff{L_0}{\rho} \right)
\displaybreak[0] \notag \\
 & \qquad + 
\left(1 -\frac{2L_0}{\rho} \right)
\left( Q_0 \ddiff{f_0}{\rho} + \diff{Q_0}{\rho} \diff{f_0}{\rho} 
\right)
\biggl. \biggr)  \Biggl. \Biggr) 
\label{apph_mathfrakA}
\end{align}  
and
\begin{equation} \label{apph_mathfrakB}
B(\rho,\, t) \equiv \Omega f_0 \left( 2 \rho^{-1}
\left(1-\frac{2L}{\rho}\right)^{-1} \di{\ }{t}L_1(\rho,\,
t)  - Q_0^{-1} \di{\ }{t}Q_1(\rho,\, t) + 2 \omega \cos\omega\,t
\right)  .
\end{equation}
The derivatives of $L_1(\rho,\,t)$ and $Q_1(\rho,\,t)$ with respect to
$\rho$ are found from the first order equations
\begin{align} 
\mbox{} &\left(1-\frac{2L_0}{\rho}\right ) Q_0 \di{\ }{\rho} L_1(\rho,\,t) 
 = -\left( \left(1-\frac{2L_0}{\rho}\right) 
Q_1(\rho,\,t)-2\rho^{-1} L_1(\rho,\,t)Q_0 \right ) \diff{L_0}{\rho} 
 \displaybreak[0] \notag \\ 
 & \qquad + \frac{1}{2} Q_0^{-1} \left( Q_0^2
\left(1-\frac{2L_0}{\rho}\right)^2 
\left ( \diff{f_0}{\rho} \right )^2 \sin \omega\,t
\right. 
+  \left(  \left (1-\frac{2L_0}{\rho}\right )^2 Q_0
Q_1(\rho,\,t) \right.
\displaybreak[0] \notag \\
 & \qquad - \Biggl. 2 \rho^{-1} Q_0^2 \left (1-\frac{2L_0}{\rho}\right)
L_1(\rho,\,t)  \Biggr) \left(  \diff{f_0}{\rho} \right )^2 
+ \Biggl. f_0^2  \sin^2 \!\omega\,t \Biggr) 
\displaybreak[0] \notag \\
 & \qquad -\frac{1}{2}  Q_0^{-2} Q_1(\rho,\,t)
\left (\frac{1}{2} \left(1-\frac{2L_0}{\rho}\right)^2
Q_0^2 \left ( \diff{f_0}{\rho} \right)^2 + \frac{1}{2} 
f_0^2\right) 
\displaybreak[0] \notag \\
 & \qquad + \frac{1}{2} l^{*2} \rho^{-2}  Q_0 f_0^2 \left(
1-\frac{2L_0}{\rho} \right) \sin\omega\,t
\displaybreak[0] \notag \\
 & \qquad  + \frac{1}{4} \left( l^{*2} \rho^{-2}\left(
1-\frac{2L_0}{\rho}\right) 
Q_1(\rho,\,t) - 2 l^{*2} \rho^{-3} L_1(\rho,\,t) Q_0 \right) f_0^2
\label{apph_L1peq}
\end{align}
and
\begin{align} 
\mbox{}
 &\left( 1- \frac{2L_0}{\rho}\right) Q_0 
\di{\ }{\rho} Q_1(\rho,\,t) =
 - \left(\left( 1- \frac{2L_0}{\rho}\right)
 Q_1(\rho,\,t) - 2 \rho^{-1} L_1(\rho,\, t) Q_0 \right)
 \diff{Q_0}{\rho}
\displaybreak[0]\notag \\
 &  \qquad  + \frac{1}{\rho - 2 L_0}  \Biggl(  \Biggr.
\left( 1- \frac{2L_0}{\rho}\right)^2 Q_0^2 
 \left(\diff{f_0}{\rho}\right)^2 \sin\omega\,t 
\displaybreak[0]\notag \\
 &  \qquad  + \frac{1}{2}   \left( 
   2 \left( 1- \frac{2L_0}{\rho}\right)^2 Q_0 
Q_1(\rho,\,t) - 4 \rho^{-1} \left( 1- \frac{2L_0}{\rho}\right)
 L_1(\rho,\, t) Q_0^2   \right)
  \left(\diff{f_0}{\rho} \right)^2 
\displaybreak[0]\notag \\
 &  \qquad  + f_0^2 \sin^2 \!\omega\,t   
  \Biggl.  \Biggr)  
 + \frac{L_1(\rho,\, t)}{\left(\rho - 2 L_0\right)^2}
   \left(
  \left( 1- \frac{2L_0}{\rho}\right)^2
 Q_0^2 \left(\diff{f_0}{\rho}\right)^2 + f_0^2 \right)   ,
\label{apph_Q1peq}
\end{align} 
respectively. The derivative of $L_1(\rho,\,t)$ with respect to $t$ is
given by the first order equation
\begin{equation} \label{apph_lteq}
\di{\ }{t}L_1(\rho,\,t) = \frac{1}{2}\omega 
\left(1-\frac{2L_0}{\rho}\right) f_0 \diff{f_0}{\rho}\cos\omega t  .
\end{equation}
The simplest of the first order equations is \eqref{apph_lteq}. It is
immediately integrable to yield
\begin{equation} \label{apph_L1equation}
L_1(\rho,\,t) = \frac{1}{2} \left(1-\frac{2L_0}{\rho}\right) f_0
\diff{f_0}{\rho}\sin\omega t  + c_1(\rho)  , 
\end{equation}
where $c_1(\rho)$ is a function of $\rho$.  It is possible to
obtain an equation for $\partial Q_1(\rho,\,t)/\partial t$ from
\eqref{apph_first_wave}. This is done by substituting
\eqref{apph_lteq} into the coefficient of $\cos\Omega\,t$ ({\em viz.}\
$B(\rho,\,t)$) and setting the expression to zero. This is
justified since $\sin\Omega\,t$ and $\cos\Omega\,t$ are independent
functions in the approximation that $A(\rho,\,t)$ and
$B(\rho,\, t)$ are slowly varying functions of
time. Solving for $\partial Q_1(\rho,\,t)/\partial t$ and integrating
yields
\begin{equation} \label{apph_Q1equation}
Q_1(\rho,\,t) = Q_0 \left( 2 + \rho^{-1} f_0 
\diff{f_0}{\rho} \right) \sin\omega t  + c_2(\rho).
\end{equation}

Up to this point the only first order field equation which has been
satisfied is \eqref{apph_lteq}. To satisfy the first order wave
equation \eqref{apph_first_wave}, the condition $A(\rho,\,t) =0$ must be
imposed. By using the first order field equations \eqref{apph_L1peq}
and \eqref{apph_Q1peq}, the $\rho$ derivatives of $L_1(\rho,\, t)$ and
$Q_1(\rho,\, t)$ found in $A(\rho,\,t)$ (Eq.~\eqref{apph_mathfrakA}) can
be eliminated. After substitution of \eqref{apph_L1equation} and
\eqref{apph_Q1equation} into \eqref{apph_mathfrakA}, $A(\rho,\,t)$ no
longer depends on the first order functions $L_1(\rho,\, t)$ and
$Q_1(\rho,\, t)$. As a result of these substitutions,
\eqref{apph_mathfrakA} takes the form\cite{fn7}
\begin{equation} \label{apph_sin_coeff}
{\cal A}(\rho) \sin\omega t + {\cal B}(c_1(\rho),\,c_2(\rho),\,\rho) =
0 , 
\end{equation}
where ${\cal A}(\rho)$ and ${\cal B}(\rho)$ depend only on $\rho$, the
unperturbed functions $f_0(\rho)$, $L_0(\rho)$, $Q_0(\rho)$ (and their
derivatives) and the two functions $c_1(\rho)$ and $c_2(\rho)$.  Note
that $c_1(\rho)$ and $c_2(\rho)$ are only found in ${\cal
B}(\rho)$. Equation \eqref{apph_sin_coeff} will be satisfied for all
$t$ only if ${\cal A}(\rho) =0$ and ${\cal B}(\rho) =0$. Since
$c_1(\rho)$ and $c_2(\rho)$ are yet to be determined, we will focus
upon the equation ${\cal A}(\rho) =0$.

The function ${\cal A}(\rho)$ is comprised of the known functions
$f_0(\rho),\,L_0(\rho)$ and $Q_0(\rho)$.  It is therefore necessary to
transform ${\cal A}(\rho)$ to the $x$ coordinate system and expand in
inverse powers of $l^{*1/3}$ as is done for the unperturbed system
\cite{Wheeler55}. The transformation is given by
Eqs.~\eqref{x_transformation}--\eqref{ell_expansion}. 
After a lengthy computation, the asymptotic expansion of ${\cal
A}(\rho) =0$ yields
\begin{equation} \label{apph_one_third}
 l^{*1/3} \left( k^{-2}(x)\left(10\lambda_0(x) - 5 + k^{2}(x)\right) +
\frac{\omega^2}{\Omega^2} \right)\phi(x) 
+ \mbox{O}\left( 1 \right) = 0  .
\end{equation}
In order for \eqref{apph_one_third} to be satisfied for large
arbitrary $l^{*}$ (in the limit $l^{*}\rightarrow \infty$), each order
of $l^{*1/3}$ must be set to zero. Since setting $\phi(x)=0$ implies
the absence of electromagnetic wave perturbations, the bracketed term
must be zero. It is known from the unperturbed system that
\cite{Wheeler55}
\begin{equation} \label{apph_lambda_k}
\lambda_0(x) = \frac{1}{2}\left(1-k^2(x)\right) .
\end{equation}
Substitution of \eqref{apph_lambda_k} in \eqref{apph_one_third} leads
to the relation
\begin{equation} \label{apph_contradiction}
\omega^2 = 4 \Omega^2
\end{equation}
which must hold in order for the field equation
\eqref{apph_first_wave} to be satisfied. However,
Eq.~\eqref{apph_contradiction} is a {\em contradiction} to the
original assumption $\omega \ll \Omega$. This result is identical to
that found for the perturbation analysis based upon
Eq.~\eqref{perturbation1}. 

\appendix{Angle Average of $T_\mu^\nu$}
%
%
%
\label{App5}
\noindent
Equations \eqref{angle_avg_tt}--\eqref{angle_avg_rt} includes
integrating over the angle $\varphi$ and dividing by the solid angle
$4\pi$, thus all that is left is evaluating the $\theta$
integrals. The $\theta$ dependence of $\T{\mu}{\nu}$ comes in three
forms
\begin{equation}
\sin^{-2}\theta \left(\Theta_l(\theta)\right)^2,\ \sin^{-2}\theta
\left(\Theta_l(\theta)_{,\theta}\right)^2 \mbox{ and }
\sin^{-2}\theta\; \Theta_l(\theta) \Theta_l(\theta)_{,\theta\theta}   , 
\end{equation}
where
\begin{equation}
\Theta_l(\theta) = C_l^0 B_l(\theta)
\end{equation}
and
\begin{equation}
B_l(\theta) \equiv \sin\theta \, \frac{\dd}{\dd\theta}\,
P_l(\cos\theta)   . 
\end{equation}
The exact integrals are evaluated below: 
\begin{equation}
\int^\pi_0 \sin^{-2}\theta \left(B_l(\theta)\right)^2 \sin\theta\,
\dd\theta = \frac{2 l (l + 1)}{2 l + 1},
\end{equation}
\begin{equation}
\int^\pi_0 \sin^{-2}\theta  
\left(B_l(\theta)_{,\theta}\right)^2 \sin\theta\,
\dd\theta = \frac{2 l^2 (l + 1)^2}{2 l + 1}, 
\end{equation}
\begin{equation}
\int^\pi_0 \sin^{-2}\theta 
B_l(\theta) B_l(\theta)_{,\theta\theta}  \sin\theta\,
\dd\theta = -\frac{2 l^3 (l + 1)}{2 l + 1} .
\end{equation}
The normalization constant for $\Theta_l(\theta)$ is found by
requiring
\begin{equation}
\int^{2\pi}_0 \int^\pi_0  
\left|\Theta_l(\theta)\right|^2    \sin\theta   \, 
\dd\theta \dd\varphi = 1     .
\end{equation}
Therefore
\begin{equation}
\left(C_l^0 \right)^2  = \frac{1}{2\pi}\left( 
 \int^\pi_0  \left(B_l(\theta)\right)^2 \sin\theta\,
\dd\theta \right)^{-1} = \frac{1}{2\pi}\left(\frac{4 l^2(l+1)^2}{(2
l-1)(2l+1)(2l+3)}  \right)^{-1}.
\end{equation}
Thus the normalization constant is
\begin{equation}
C_l^0 = \left(\frac{(2l-1)(2l+1)(2l+3)}{8\pi l^2(l+1)^2} \right)^{1/2}.
\end{equation}

\newpage
\nonumsection{References}


\begin{thebibliography}{00}
\bibitem{fn1} To perform the numerical integration, a Fehlberg
fourth-fifth order Runge--Kutta method from the MAPLE V R4 program
library is used.
\bibitem{fn2} It is convenient to use the functions $ \phi,\, u$ in
place of $f_1,\, f_2$ for the remainder of this section. Recall
$\phi=f_1,\, u=f_2$ from Eq.~\eqref{phi-f1-transformation}.
\bibitem{fn3} Existence of the eigenvalue was derived independently in
Ref.~\citen{AndersonBrill} using an alternate method. However,
stability aspects were not discussed in Ref.~\citen{AndersonBrill}.
\bibitem{fn4} The active modes are characterized by a sequence of
parameter values (a family of modes) for which the effective
stress-energy is concentrated in the same spatial region.
\bibitem{fn5} These are known as numerical solutions to the system
\eqref{geonwave}--\eqref{geonj} with initial conditions
\eqref{eq4.40}--\eqref{eq4.43} in the high angular momentum
approximation.
\bibitem{fn7} Explicit forms of ${\cal A}(\rho)$ and ${\cal B}(\rho)$
will not be given due to their extreme length.
\end{thebibliography}
\end{document}